\documentclass[12pt,preprint]{aastex}
\begin{document}

\title{Determination of the intrinsic Luminosity Time Correlation in the X-ray Afterglows of GRBs}

\author{Maria Giovanna Dainotti\altaffilmark{1,2}, Vahe' Petrosian \altaffilmark{1}, Jack Singal \altaffilmark{1,3},
 Michal Ostrowski\altaffilmark{1}}
 
\altaffiltext{1}{Department of Physics \& Astronomy, Stanford University, Via Pueblo Mall 382, Stanford CA 94305-4060, E-mail: vahep@stanford.edu;jacks@slac.stanford.edu;mdainott@stanford.edu}

\altaffiltext{2}{Obserwatorium Astronomiczne, Uniwersytet Jagiello\'nski, ul. Orla 171, 31-501 Krak{\'o}w, Poland, E-mails: dainotti@oa.uj.edu.pl,mio@oa.uj.edu.pl}

\altaffiltext{3}{Physics Department, University of Richmond, 28 Westhampton Way, Richmond, VA 23173,jacks@slac.stanford.edu}
\date{\today}
\begin{abstract}
Gamma-ray bursts (GRBs), which have been observed up to redshifts $z \approx 9.5$ can be good probes of the early universe and have the potential of testing cosmological models. The analysis by Dainotti of GRB Swift afterglow lightcurves with known redshifts and definite X-ray plateau shows an anti-correlation between the \underline{rest frame} time when the plateau ends (the plateau end time) and the calculated luminosity at that time (or approximately an anti-correlation between plateau duration and luminosity). We present here an update of this correlation with a larger data sample of 101 GRBs with good lightcurves. 
Since some of this correlation could result from the redshift dependences of these intrinsic parameters, namely their cosmological evolution we use the Efron-Petrosian method to reveal the intrinsic nature of this correlation. We find that a substantial part of the correlation is intrinsic and describe how we recover it and how this can be used to constrain physical models of the plateau emission, whose origin is still unknown. The present result could help clarifing the debated issue about the nature of the plateau emission.
\end{abstract}

 
\maketitle
\section{Introduction}
GRBs are the farthest sources, seen up to redshift $z=9.46$ \citep{Cucchiara2011}, and if emitting isotropically they are also the most powerful, (with $E_{iso} \leq 10^{54}$  erg  s$^{-1}$), objects in the Universe. In spite of the great diversity of their prompt emission lightcurves and their broad range spanning over 7 orders of magnitude of $E_{iso}$, some common features have been identified from investigation of their afterglow light curves.
A crucial breakthrough in this field has been the observation of GRBs by the \textit{Swift} satellite which provides a rapid follow-up of the afterglows in several wavelengths revealing a more complex behavior of the X-ray lightcurves than a broken power law generally observed before \cite{OB06,Sak07}. The {\it Swift} afterglow lightcurves manifest several segments. The second segment, when it is flat, is called the plateau emission.
A significant step forward in determining common features in the afterglow lightcurves was made by fitting them with an analytical expression \cite{W07}, called hereafter W07. 

This provides the opportunity to look for universal features that could provide a redshift independent measure of the distance, as in studies of correlations between GRB isotropic energy and peak photon energy of the $\nu F_{\nu}$ spectrum, $ E_{iso}$-$E_{peak}$, \cite{Lloyd1999,amati09}, the beamed total energy $E_{\gamma}$\,-\,$E_{peak}$ \cite{G04,Ghirlanda06}, $L$\,-\,$V$ luminosity-Variability, \cite{N00,FRR00}, $L$\,-\,$E_{peak}$ \cite{Yonekotu04} and possibly others \cite{S03}. Impacts of detector thresholds on cosmological standard candles have also been considered \cite{Shahmoradi09,Petrosian1998,Petrosian1999,Petrosian2002,Cabrera2007}.
Unfortunately, because of large dispersion \cite{Butler09,Yu09} and absence of good calibration none of these correlations allow the use of GRBs as `standard candles' as has been done e.g. with type Ia Supernovae. 

Dainotti et al. (2008, 2010), using the W07 phenomenological law for the lightcurves of long GRBs, discovered a formal anti-correlation  between the X-ray luminosity at the end of the plateau $L_X$ and the rest frame plateau end- time, $T^{*}_a=T^{obs}_a/(1+z)$, (hereafter LT), described as :

\begin{equation}
\log L_X = \log a + b \log T^*_{a}, 
\label{feq}
\end{equation}
where $T^{*}_a$ is in seconds and $L_X$ is in erg/s. The normalization and the slope parameters $a$ and $b$ are constants obtained by the D'Agostini fitting method \citep{Dago05}.  
Dainotti et al. 2011a attempted to use the LT correlation as possible redshift estimator, but the paucity of the data and the scatter prevents from a definite conclusion at least for a sample of 62 GRBs. In addition, a further step to better understand the role of the plateau emission has been made with the discovery of new significant correlations between $L_X$, and the mean luminosities of the prompt emission, $<L_{\gamma,prompt}>$ \citep{Dainotti2011b}.
 
The LT anticorrelation is also a useful test for theoretical models such as the accretion models, \citep{Cannizzo09,Cannizzo11}, the magnetar models \citep{Dall'Osso,Bernardini2011,Bernardini2012,Rowlinson2010,Rowlinson2013}, the prior emission model \citep{Yamazaki09}, the unified GRB and AGN model \citep{Nemmen2012} and the fireshell model \citep{Izzo2012}. Furthermore, it has been recovered within also other observational correlations \citep{Ghisellini2008,Sultana2012,Qi2012}. Finally, it has been applied as a cosmological tool \citep{Cardone09,Cardone2010,Postnikov2013}.
Here, we study an updated sample of 101 GRBs and we investigate whether the LT correlation is intrinsic or induced by cosmological evolution of $L_X$ and $T^{*}_a$, and/or observational biases due to the instrumental threshold. This step is necessary to cast light on the nature of the plateau emission, to provide further constraints on the theoretical models, and possibly to assess the use of the LT correlation as a model discriminator.
In section \ref{Data} we describe the data and the results from correlation test carried using the \underline{raw} data. In section \ref{intrinsic correlations} we use the EP method to determine the \underline{intrinsic} correlation between $L_X$ and $T^*_{a}$. In section \ref{density and luminosity} the cumulative density and luminosity are defined and derived. This is followed by a discussion section.

\section{Lightcurve Data and raw correlations}\label{Data}

We have analyzed the sample of all GRB X-ray afterglows with known redshifts detected by {\it Swift} from January 2005 up to May 2011, for which the light curves include early X-ray data and therefore can be fitted by the W07 model. Willingale 
proposed a functional form for $f(t)$ :

\begin{equation}
f(t) = \left \{
\begin{array}{ll}
\displaystyle{F_i \exp{\left ( \alpha_i \left( 1 - \frac{t}{T_i} \right) \right )} \exp{\left (
- \frac{t_i}{t} \right )}} & {\rm for} \ \ t < T_i \\
~ & ~ \\
\displaystyle{F_i \left ( \frac{t}{T_i} \right )^{-\alpha_i}
\exp{\left ( - \frac{t_i}{t} \right )}} & {\rm for} \ \ t \ge T_i \\
\end{array}
\right .
\label{eq: fc}
\end{equation}

for both the prompt (the index `i=\textit{p}') $\gamma$\,-\,ray and initial X -ray decay and for the afterglow ( ``i=\textit{a}") modeled so that the complete lightcurve $f_{tot}(t) = f_p(t) + f_a(t)$ contains two sets of four parameters $(T_{i},F_{i},\alpha_i,t_i)$. The transition from the exponential to the power law occurs at the point $(T_{i},F_{i}e^{-t_i/T_i})$ where the two functional sections have the same value and gradient. The parameter $\alpha_{i}$ is the temporal power law decay index and the time $t_{i}$ is the initial rise timescale.

In previous papers W07, Dainotti et al. (2008,2010) fitted the {\it Swift} Burst Alert Telescope (BAT)+ X-Ray Telescope (XRT) lightcurves of GRBs were fitted to Eq. (\ref{eq: fc}) assuming that the rise time of the afterglow started at the time of the beginning of the decay phase of the prompt emission, $T_p$, namely $t_a=T_p$. In this paper, we search for an independent measure of the parameters of the afterglow, thus we leave $t_a$ to be a free parameter. In the majority of the cases we have $t_a \geq 0$.
We use the redshifts available in the literature \citep{X09}, in Greiner web page http://www.mpe.mpg.de/~jcg/grbgen.html, and in the Circulars Notice arxive (GCN). (We exclude GRBs with uncertain redshift measurement.) The complete set of GRBs with definite known redshift till May 2011 is $> 120$, but not all GRBs show a well defined plateau emission.
 {\bf The fitting procedure fails either when it gives unreasonable values, or when the determination of confidence interval in 1 $\sigma$ doesn't fulfill the Avni 1976 prescriptions, for more details see http://heasarc.nasa.gov/xanadu/xspec/manual/XspecSpectralFitting.html.
 The latter prescriptions require for a proper evaluation of the error bars the computation in the 1 $\sigma$ confidence interval for every parameter varying the parameter value until the $\chi^2$ increases by $3.5$ above the minimum (or the best-fit) value, because we are in a tree-parameter space. These rules define the amount that the $\chi^2$ is allowed to increase, which depends on the confidence level one requires, and on the number of parameters whose confidence space is being calculated.}
 
The source rest-frame luminosity in the {\it Swift} XRT bandpass, $(E_{min}, E_{max})=(0.3,10)$ keV at time $T_a$, is computed from the Equation:

\begin{equation}
L_X (E_{min},E_{max},T_a)= 4 \pi D_L^2(z) \, F_X (E_{min},E_{max},T_a) \times \textit{K},
\label{eq: lx}
\end{equation}

where $D_L(z)$ is the GRB luminosity distance \footnote{We assume a $\Lambda$CDM flat cosmological model with $\Omega_M = 0.291$ and $H_0 = 71  {\rm Km s}^{-1} {\rm Mpc}^{-1}$}, $F_X$ is the measured X-ray energy flux and $\textit{K}= (1+z)^{-1 +\beta_{a}}$ is the so called \textit{K}-correction for X-ray power law index $\beta_{a}$ \cite{Evans2009,Dainotti2010}. {\bf The error bars on the normalization parameter and the slope quoted of the $L_X$ and $T^*_{a}$ are computed with the method of D’Agostini 
(2005), which is a suitable method where the errors on both variables are comparable (which is the case here)
and it is not possible to decide which one is the independent variable to be used in the $\chi^2$ fitting analysis.
Moreover, the relation $L_X = a T_{a}^b$ may be affected by an intrinsic
scatter $\sigma_{int}$ of unknown nature that has to be taken into
account. Thus, to determine the parameters $(a, b,\sigma_{int})$, we follow the \cite{Dago05} Bayesian approach
 and maximize the likelihood function
${\cal{L}}(a, b, \sigma_{int}) = \exp{[-L(a, b, \sigma_{int})]}$
where:

\begin{eqnarray}
L(a, b, \sigma_{int}) & = &
\frac{1}{2} \sum{\ln{(\sigma_{int}^2 + \sigma_{y_i}^2 + b^2
\sigma_{x_i}^2)}} \nonumber \\
~ & + & \frac{1}{2} \sum{\frac{(y_i - a - b x_i)^2}{\sigma_{int}^2 + \sigma_{y_i}^2 + b^2
\sigma_{x_i}^2},}
\label{eq: deflike}
\end{eqnarray}
 $(x_i, y_i) = (\log{L_X}, \log{T_a})$ and the sum is over the
${\cal{N}}$ objects in the sample. Note that, actually, this maximization
is performed in the two parameter space $(b, \sigma_{int})$ since $a$ may
be estimated analytically as\,:

\begin{equation}
a = \left [ \sum{\frac{y_i - b x_i}{\sigma_{int}^2 + \sigma_{y_i}^2 + b^2
\sigma_{x_i}^2}} \right ] \left [\sum{\frac{1}{\sigma_{int}^2 + \sigma_{y_i}^2 + b^2
\sigma_{x_i}^2}} \right ]^{-1}
\label{eq: calca}
\end{equation}
so that we will not consider it anymore as a fit parameter.

 \footnote{We pointed out here that since this method takes into account the hidden errors thus gives greater error estimates than
the ones obtained with the Marquardt Levemberg algorithm (Marquardt 1963).} 
Initially, we had a sample of 116 GRBs with firm redshift including 11 IC, and with the evaluation of the observables $T_a$, $F_a$, $\alpha_a$ but not for all of them we were able to fulfill the Avni prescriptions mentioned above. Among the 116 GRBs, 104 had the proper evaluation of the error measurements, but 3 of them had an error energy parameter $\sigma_{E} \equiv \sqrt{\sigma_{L^*_{X}}^2 + \sigma_{T^*_a}^2} > 1$, (for definition about this parameter and its use, see Dainotti et al. 2011b) therefore we discarded, because such values of the errors have no physical meaning. We thus have a sample of 101 GRBs.
To ensure that the inclusion of the IC does not introduce biases in the evaluation of the power slope for the $L_X$-$T^*_a$ correlation for long GRBs, we checked the slopes of the sample with and without the 8 IC bursts.
The two power slopes are compatible within 1 $\sigma$. We pointed out that in a previous paper \citep{Dainotti2010} we did not introduce the IC bursts because these represented more than $14 \%$ of the sample, while in the current sample they represents only $8 \%$. For the whole sample without the IC we found the power law slope $b=-1.27 \pm _{-0.26}^{+0.18}$, while for the whole sample $b=-1.32 \pm _{-0.17}^{+0.18}$.} The Spearman correlation coefficient for the larger sample ($\rho=-0.74$) is higher than $\rho=-0.68$ obtained for a subsample of 66 long duration GRBs analyzed in Dainotti et al. 2010. The probability of the correlation (of the 101 long GRBs) occurring by chance within an uncorrelated sample is $P \approx 10^{-18}$ \cite{Bevington}. 

Figure \ref{fig1}, left panel, shows the $L_X$-$T^*_a$ distribution of 101 GRBs with $0.08 \leq z \leq 9.4$ and includes afterglows of 93 long and 8 short bursts with extended emission \citep{nb2010}, called the Intermediate class (IC), see Table 1 \footnote{for a complete table of the fitting parameters see http://www.oa.uj.edu.pl/M.Dainotti}

\begin{table*}
\caption{Fitting parameters, the first column is the GRB identification number, the second, z, the redshift, the third, Fx, the X - ray observed flux, the fourth its error, $dFx$,  the fifth, $beta_a$, the spectral index, the sixth, the error on the spectral index, $dbeta_a$, the seventh is log Ta*, the logarithm of the characteristic rest frame time, the eighth is the error on logTa, $dlogTa$, the ninth, log Lx, the logarithm of the X - ray source luminosity at $T_a$, the tenth is the error on log Lx, $dlogLx$, the last column is the class, namely indication of the GRB type, long, IC (intermediate class).}\label{Table 1}

\begin{tabular}{|l|l|l|l|l|l|l|l|l|l|l|}
\hline
$GRB$ &	 $z$ &  $Fx$ & $dFx$ & $beta_a$ & $dbeta_a$ &$log Ta*$ & $dlogTa$ & $log Lx$ & $dlog Lx$ & $class$    \\ \hline
\hline
 50315 &	1.949	& 1.16e-11 & 1.56e-12 &	1.47 & 1.23 &	3.97 & 0.09	&47.49 & 0.56&    long \\
50318  & 1.44 &  	1.0e-8  & 1.41e-9	& 0.93 & 0.18 &    1.62 & 0.59 &    50.09 & 0.62 &  long\\
50401	& 2.9	& 5.41e-11 & 1.41e-11 &	0.87 & 0.23 &	3.19 & 0.04 &	48.58 & 0.12 &    long\\
050416A	& 0.6535 &	2.82e-11 & 3.82e-12 &	1.16 & 0.32	& 2.86 & 0.09 &	46.70 & 0.11 &    long\\
050505 &  4.27 &  4.93e-12 &3.84e-12 &   1.09 & 0.04  & 3.67 & 0.09 &48.02 & 0.34 &   long\\
050525A &	0.606 &	2.92e-9 & 6.81e-10 &	1.04 & 0.15 &	2.29 & 0.10 &	48.84 & 0.11 & long \\
050603	& 2.82 &	1.10e-12 & 6.64e-13 &	0.91 & 0.10	& 4.25 & 0.25 & 46.82 & 0.27	& IC \\
50730	&3.97	&2.58e-11 & 1.55e-12 &	0.54 & 0.05	& 3.46 & 0.01 &	48.58 & 0.04 &    long \\
50802 &	1.71 &	2.20e-11 & 1.49e-12 &	0.82 & 0.08 &	3.39 & 0.02 &	47.63 & 0.04 &    long \\
050820A &	2.612	& 6.28e-11 & 5.12e-12	& 0.91 & 0.10 &	3.40 & 0.03 &	48.53 & 0.06 &    long \\
50824	& 0.83 &	5.37e-13 & 1.10e-13 &	0.95 & 0.14 &	4.91 & 0.15 &	45.24 & 0.10 &    long \\
050904	&6.29&	5.79e-12 & 6.16e-13 &	0.61 & 0.02 &	3.15 & 0.40 &	48.09 & 0.46 &	 long\\
050922C	& 2.198 &	8.54e-12 & 2.29e-12 &	0.92 & 0.24 &	3.38 & 0.09	& 47.48 & 0.16 &    long\\
051016B	& 0.9364 &	3.22e-12 & 5.60e-13 &	0.83 & 0.15 &	3.83 & 0.11 &	46.14 & 0.09 &    long\\
051109A	&2.35 &	2.51e-11 & 7.74e-12 &	0.93 & 0.02 &	3.40 & 0.11	& 48.01 & 0.13 &    long\\
051221A	& 0.5465 &	7.91e-13 & 1.06e-13 &	0.95 & 0.18 &	4.47 & 0.07 &	44.96 & 0.08 &     IC\\
60108	&2.03	&1.69e-12 & 2.99e-13&	1.00 & 0.24&	3.80 & 0.09&	47.17 & 0.14  &  long\\
60115	&3.53&	3.51e-12 & 6.62e-13&	0.96 &  0.21	&3.06 & 0.11	&47.59 & 0.14&    long\\
60124	&2.297&	4.02e-11& 3.25e-12&	0.97 & 0.14&	3.75 & 0.03&	48.20 & 0.07&    long\\ 
60206	&4.05	&5.69e-11 & 1.61e-11&	1.29 & 0.59&	3.12 & 0.08	&48.95 & 0.35 &   long\\
060210&	3.91&	4.84e-12 & 2.65e-12&	1.05 & 0.04&	3.77 & 0.22&	47.90 & 0.24&	long\\
60218	&0.0331	&1.32e-12 & 5.34e-13&	3.51 & 0.45	&5.29 &0.13&	42.52 & 0.18&    long\\
060223A&	4.41&	1.14e-11 & 5.98e-12&	1.02 & 0.12&	1.99 & 0.22&	48.37 & 0.24&	long\\
60418	&1.49	&1.47e-10 & 2.17e-11&	1.04 & 0.22&	2.68 &0.07&	48.30 &0.11&    long\\
060502A	&1.51&	5.79e-12 & 5.86e-13	&1.04 & 0.11& 3.94 &0.08	&46.91 &0.06&     IC\\
060510B &4.9& 3.51e-13 & 3.96e-14 & 1.57 &0.12 & 3.78 &0.48 & 47.39 &0.5 & long  \\
60512	&2.1& 	1.60e-12 &5.69e-13 &	1.08 &0.28	& 3.31 &0.21 &	46.75 &	0.20 &    long\\
60522	&5.11&	1.88e-12 & 5.80e-13	&1.14 &0.28	&3.17 &0.14 &	47.70 &0.21 &   long\\
60526	&3.21 &	4.21e-12 &7.22e-13&	0.95 &0.11&	3.27 &0.10	&47.57 &0.09&    long\\
60604	&2.68 &	2.31e-12 &2.92e-13	&1.08 &0.10&	3.87 &0.06&	47.13 &0.07   & long\\
60605&	3.8	&6.48e-12 &1.03e-13&	1.03 &0.11	&3.32 &0.05&	47.94 &0.09&    long\\
060607A&	3.082&	4.17e-12 &5.29e-13&	0.57 &0.06	&3.77 &0.02&	47.53 &0.06 &  long\\
60614	&0.125&	1.54e-12 &2.05e-13&	0.88 &0.05&	5.01 &0.04	&43.79 &0.06&    IC\\
60707	&3.43 &	3.74e-12 &1.40e-13&	1.34 &0.18	&3.81 &0.16&	47.59 &0.19&    long\\
60714	&2.71 &	1.71e-11 &1.52e-12	&0.90 &0.15&	3.07 &0.05&	48.01 &0.09&    long\\
60729	&0.54&	7.97e-12 &2.58e-13&	1.03 &0.04&	4.88 &0.02&	46.95 &0.02&    long\\
60814	&0.84& 	2.75e-11 &2.92e-12&	1.10 &0.11&	3.71 &0.04&	46.96 &0.06&    long\\

\hline
\end{tabular}
\end{table*}

However, as mentioned above, because both $L_X$ and $T^*_a$ depend on redshift ($L_X$ increasing and $T^*_a$ decresing with $z$) and the sample covers a broad redshift range all or part of the anticorrelation might be induced by these dependencies. It is therefore important to determine the extent of this effect and determine the true or intrinsic correlation.
In addition any cosmological evolution in $L_X$ and/or $T^*_a$ will affect the degree of the observed anti-correlation.
{\bf Fig.\ref{fig1}, central panel, shows the colour coded fitted lines. The distribution of the subsamples presents different power law slopes when we divide the whole sample into 5 redshift bins (see Dainotti et al. 2011 for a comparison with a smaller sample) thus having 20 GRBs in each subsample. The objects in different bins exhibit some separation into different regions of the $L_X$-$T^*_a$ plane. The results are shown in fig \ref{fig1} (central) with the fitted lines. In the right panel of Fig. \ref{fig1} we show the power slope of the redshift bins with the mean values of the redshif bins. 

\begin{figure}
\includegraphics[width=0.33\hsize,angle=0,clip]{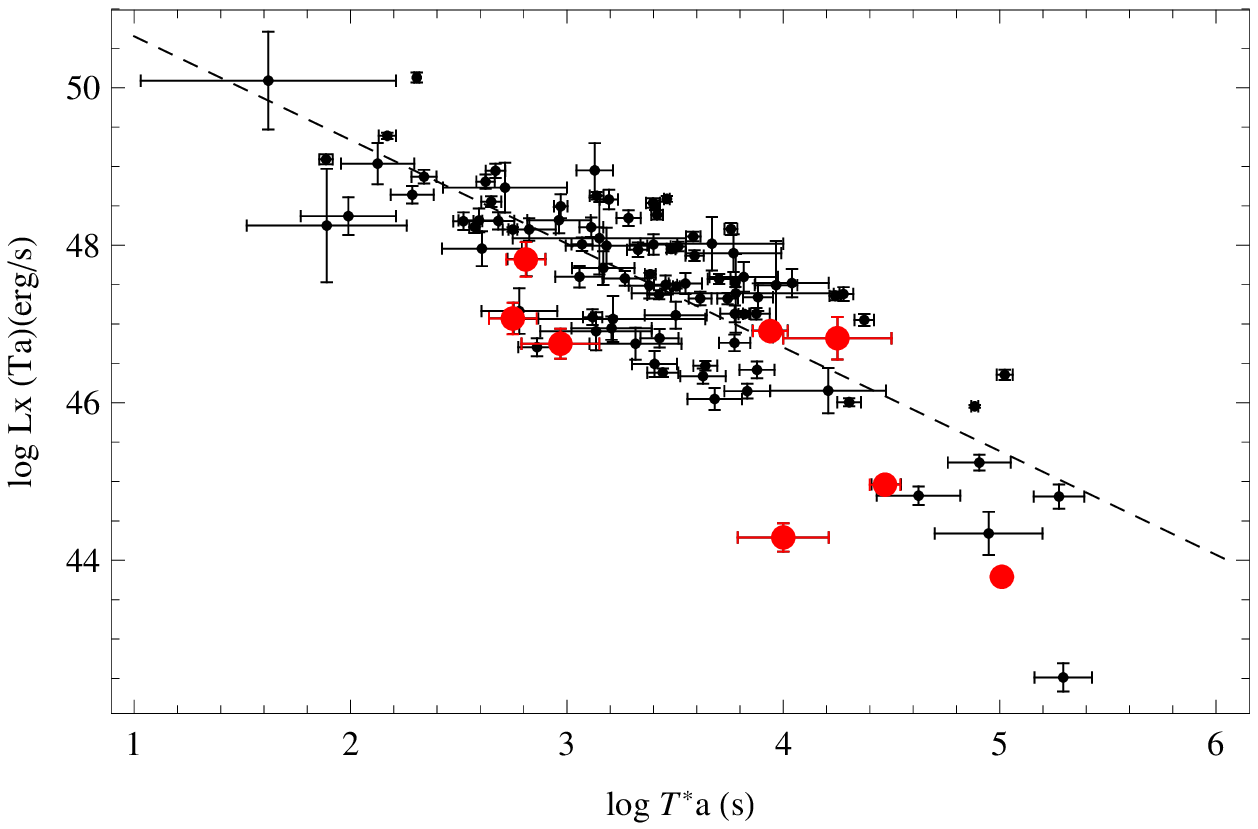}
\includegraphics[width=0.33\hsize,angle=0,clip]{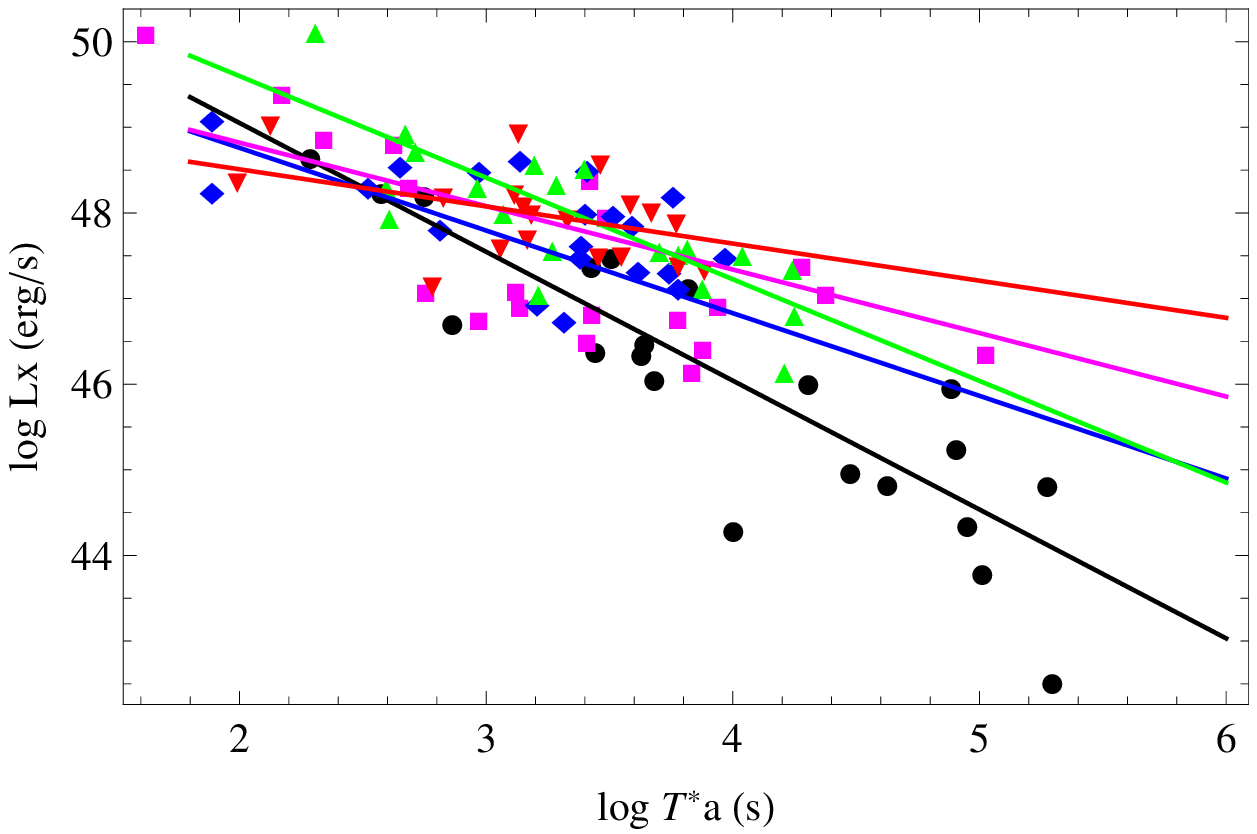}
\includegraphics[width=0.33\hsize,angle=0,clip]{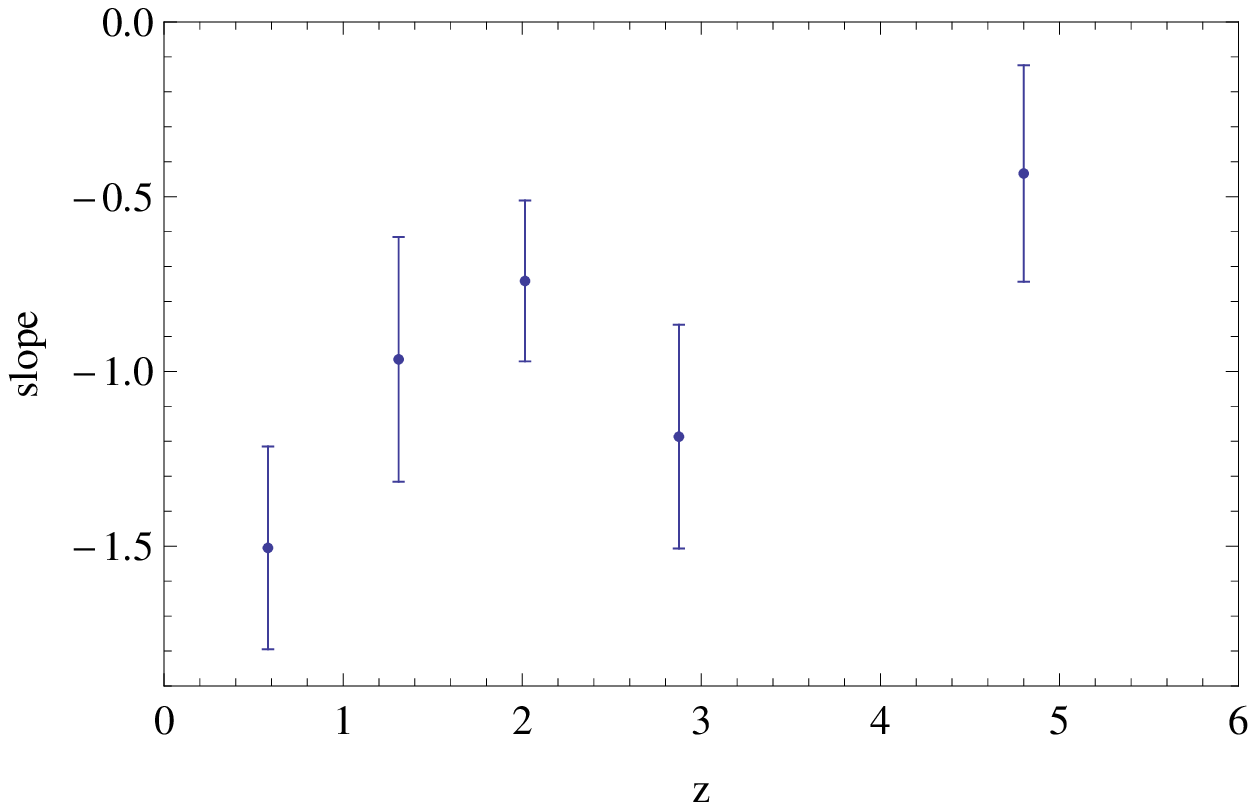}
\caption{{\bf Left Panel} $L_X$ vs $T^*_a$ distribution for the sample of 101 GRB afterglows with the fitted correlation shown by the dashed line. The red points are the IC bursts. {\bf Central Panel}: The same distribution divided in 5 equipopulated redshift bins shown by different colours: black for $z < 0.89$, magenta for $0.89 \leq z \leq 1.68$, blue for $1.68 < z \leq 2.45$, green $2.45 < z \leq 3.45$, red for $ z \geq 1.76$. Solid lines shows the fitted correlations. {\bf Right panel} The variation of the power law slope (and its error range) vith the mean value of the redshift bins.}
\label{fig1}
\end{figure}

\begin{figure}
\includegraphics[width=0.5\hsize,angle=0,clip]{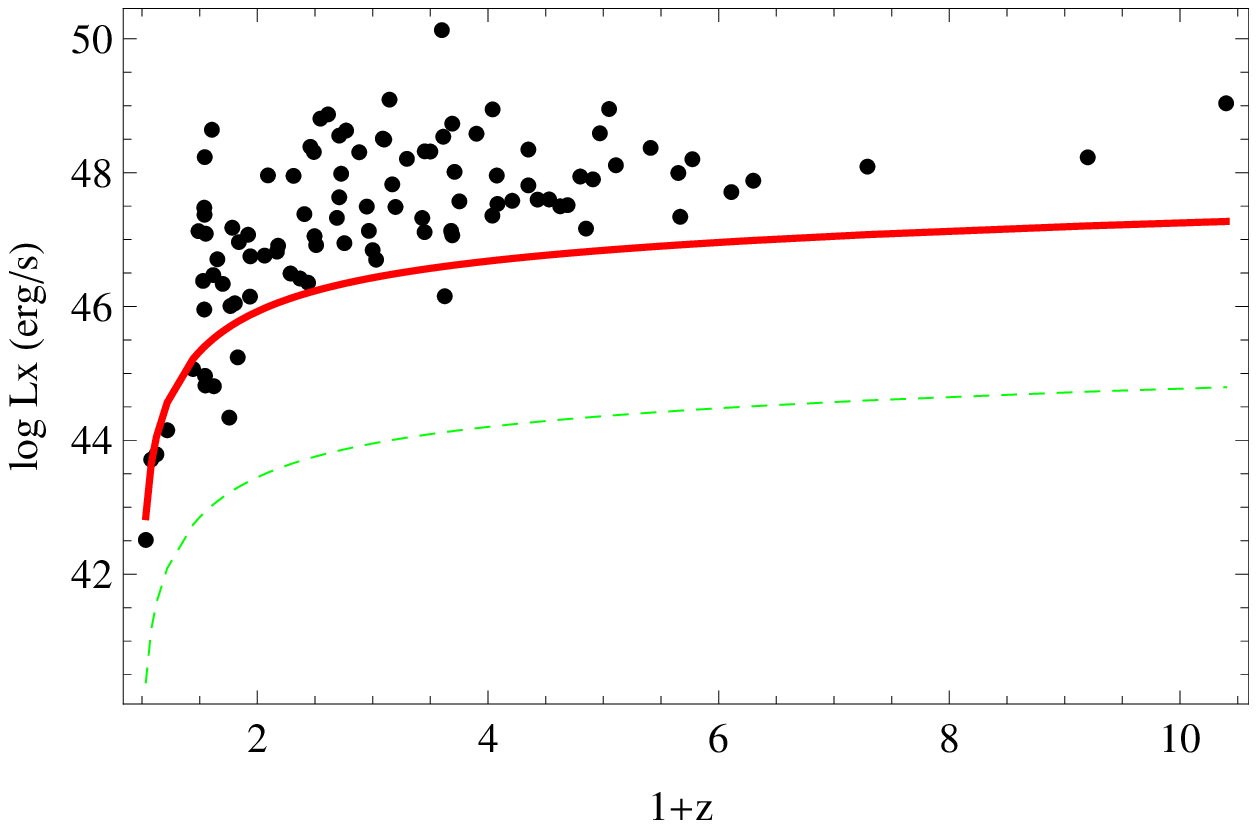}
\includegraphics[width=0.5\hsize,angle=0,clip]{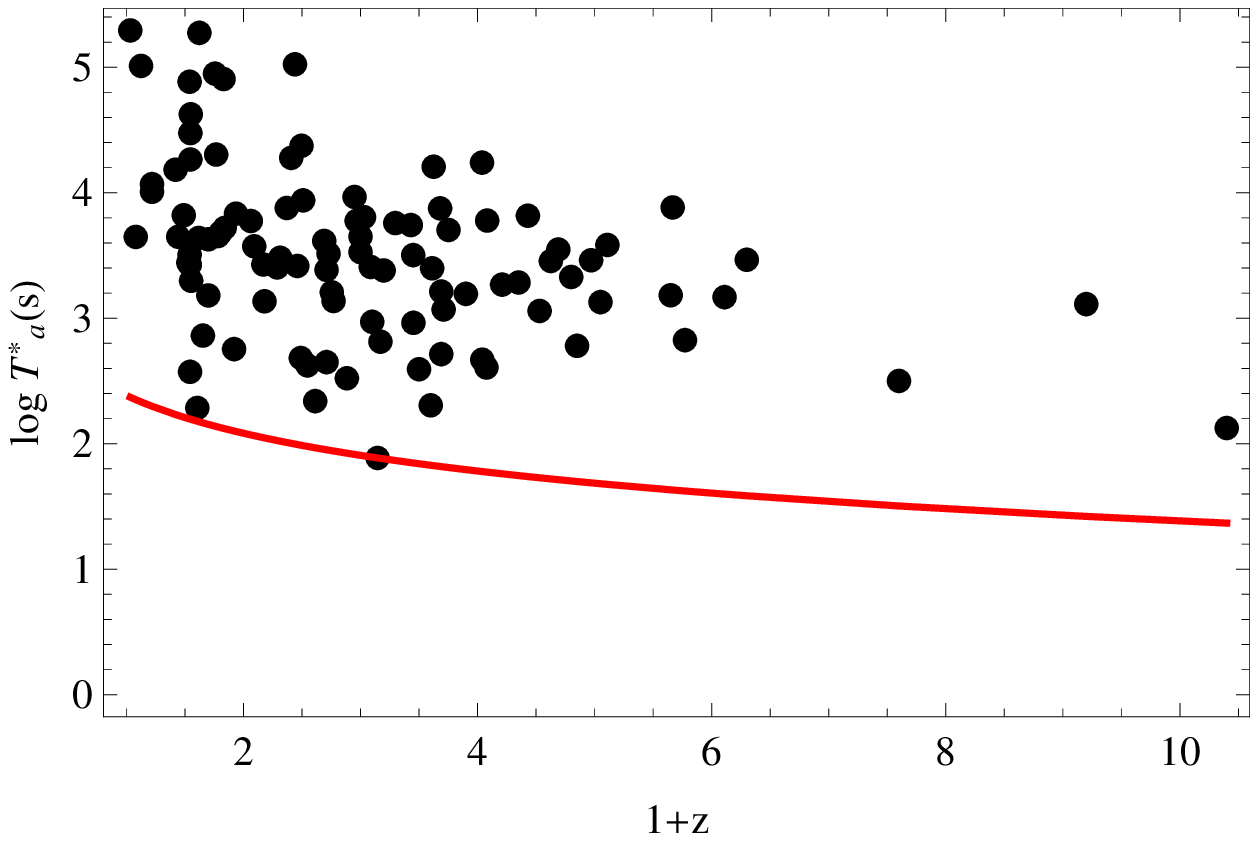}
\caption{{\bf Left Panel}: The bivariate distribution of $L_X$ and redshift with two different flux limits. The instrumental XRT flux limit,  $1.0 \times 10^{-14}$ erg cm$^{-2}$ (dashed green line) is too low to be representative of the flux limit, $1.5 \times 10^{-12}$ erg cm$^{-2}$ (solid red line) better represents the limit of the sample. 
{\bf Right panel}: The bivariate distribution of the rest frame time $T^*_a$ and the redshift. The chosen limiting value of the observed end-time of the plateau in the sample, $T_{a,lim}= 242$ s. The red line is the limiting rest frame time, $T_{a,{\rm lim}}/(1+z)$.}
\label{fig2}
\end{figure}

\begin{figure}
\includegraphics[width=0.51\hsize,angle=0,clip]{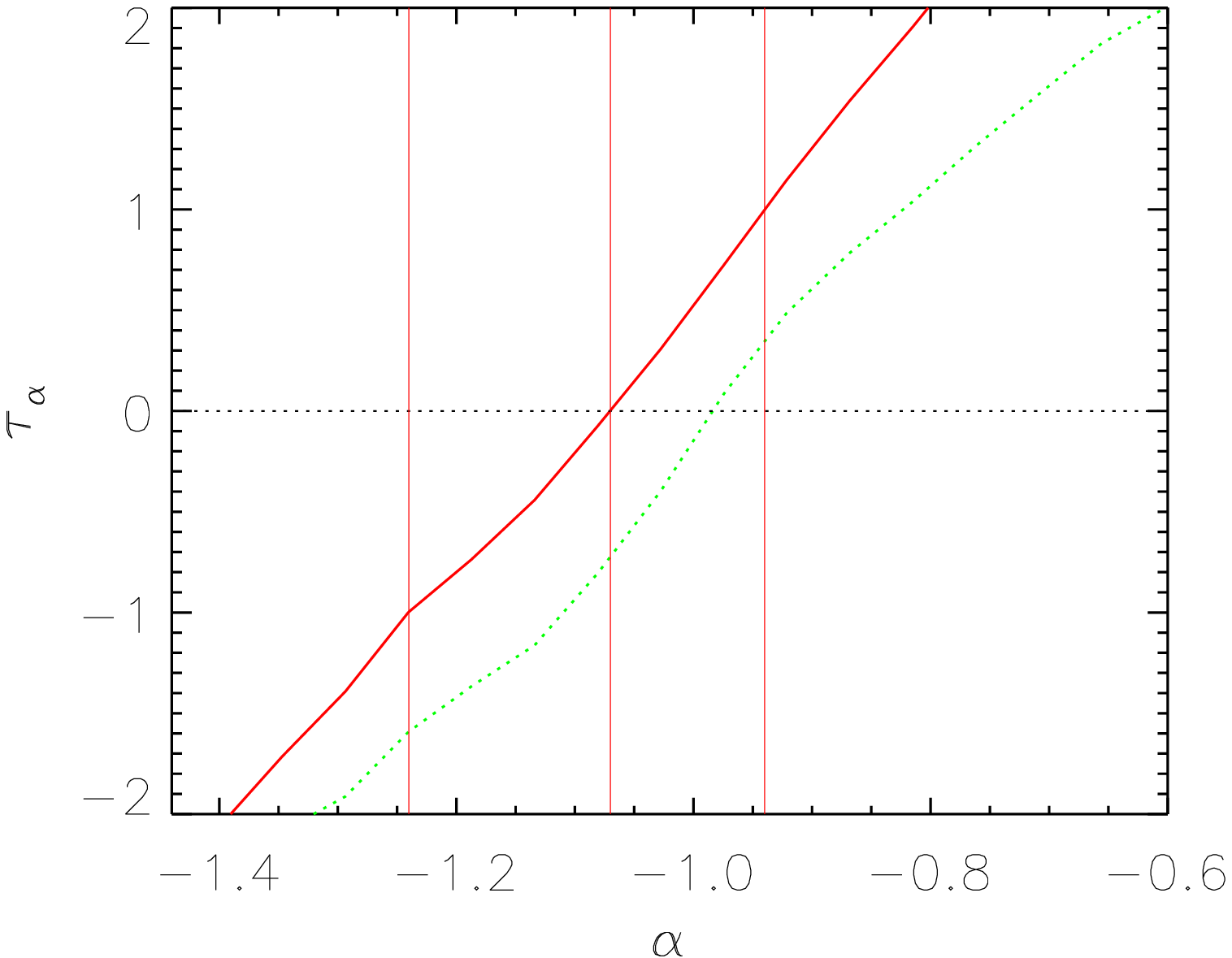}
\includegraphics[width=0.51\hsize,angle=0,clip]{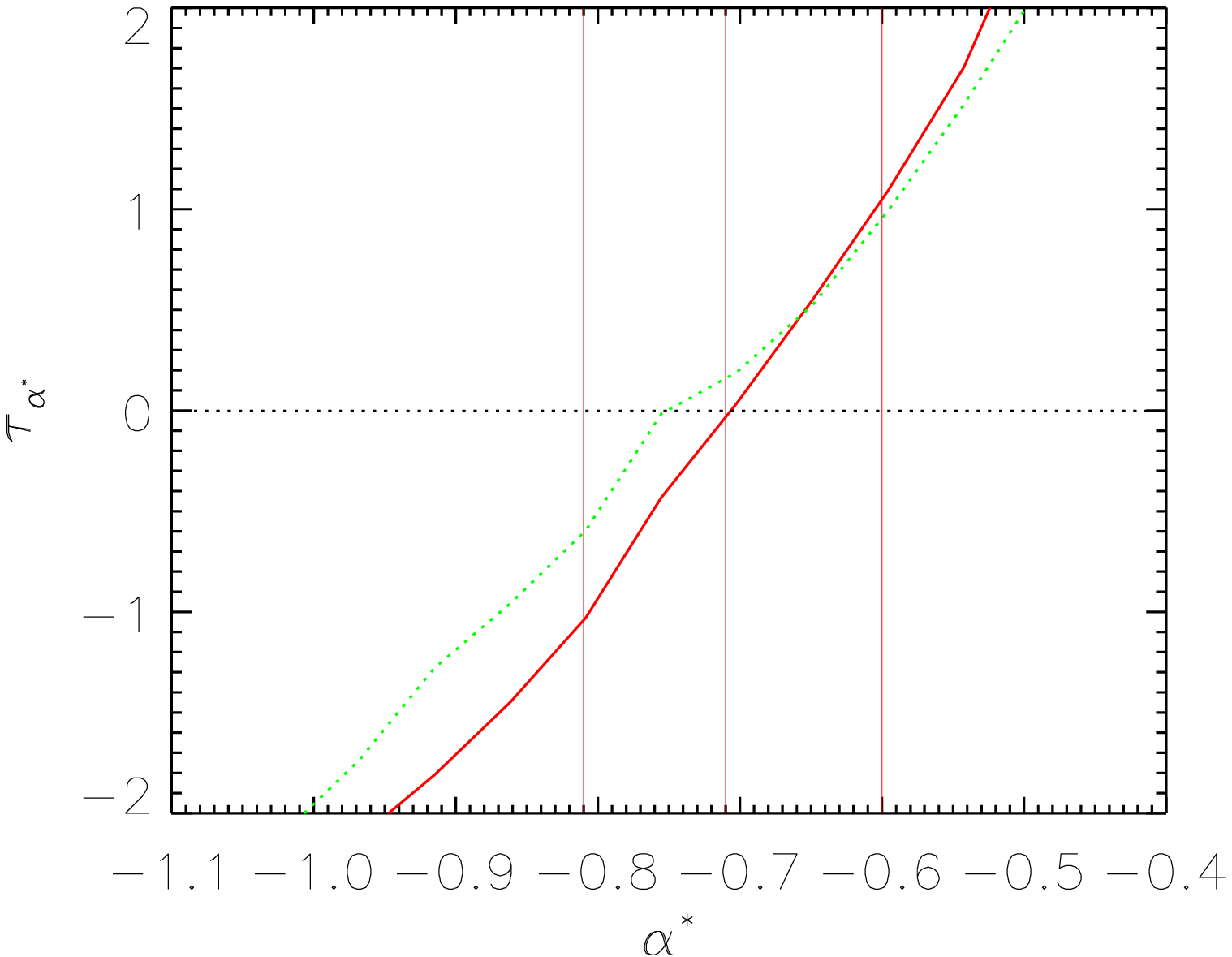}
\caption{{\bf Left Panel}: Test statistic $\tau$ vs. $\alpha$, the slope of the LT correlation. {\bf Right Panel}: Test statistic $\tau$ vs. $\alpha^{*}=1/\alpha$, the power slope of the reciprocol of the LT correlation defined by Eq. \ref{eq:alphareciprocol}. The vertical lines show the best value $\tau=0$, and the 1$\sigma$ range for $| \tau | \leq 1$ for $\alpha$ and $\alpha^{*}$. Note that we expect $\alpha=1/\alpha^{*}$ which is the case within 1$\sigma$ showing the consistency of our results. The $\tau$ values for the earlier sample of 53 GRBs are also shown by (green) dotted lines. This is also consistent with the result from the current sample of 101 GRBs.}
\label{Fig3}
\end{figure}

As evident for each bin, we again found an anticorrelation similar to the whole sample, but the mean values of the slopes are larger (smaller in absolute values) indicating flatter relations, except for the first redshift bin, than for the whole sample. As shown in the right panel of Fig. \ref{fig1} (left) there is some indication that the slope steepens for higher redshifts.} This is the first indication that some of the anticorrelation may be induced by the above mentioned effects \footnote{Note also as a result the intercept or normalization parameters a for the individual bins are smaller than the sample as a whole}.
However, in all cases these differences are all less then $3 \sigma$. 
We expect the correlation slope be closer to the one of the subsamples than the whole sample, because each subsample has a smaller redshift range $\delta_z$, which decreases the effect of the redshift dependence and/or redshift evolution. 
In addition, this test disfavors a strong redshift evolution in the correlation. In the next section we give a more quantitative analysis of these results with the Efron \& Petrosian (EP) method \citep{Efron1992} which is able to determine the intrinsic correlation among variables in a truncated bivariate distribution. 

\section{Determination of intrinsic correlations}\label{intrinsic correlations}

The first important step for determining the distribution of true correlations among
the variables is quantification of the biases introduced by the observational and sample selection effects.
In the case under study the selection effect or bias that distorts the statistical correlations are the flux limit and the temporal resolution of the instrument.
To account for these effects we apply the Efron \& Petrosian technique, already successfully applied for GRBs \citep{Petrosian2009,Lloyd2000,Kocevski2006}. Other methodologies to treat selection biases have also been investigated \citep{Collazzi2008}.

{\bf The EP method reveals the intrinsic correlation because the method is specifically designed to overcome the biases resulting from incomplete data. Moreover, it identifies and removes also the redshift evolution present in both variables, time and luminosity.}

The EP method uses a modified version of the Kendall $\tau$ statistic to test the independence of variables in a truncated data.
Instead of calculating the ranks $R_{i}$ of each data points among all observed objects, which is normally done for an untruncated data, the rank of each data point is determined among its ``associated sets" which include all objects that could have been observed given the observational limits. A full discussion of the method is provided in the literature \cite{Singal2011} and references cited therein.

{\bf Here we give a brief summary of the algebra involved in the EP method. This method uses the Kendall rank test to determine the best-fit values of parameters describing the correlation functions using  the test statistic 

\begin{equation}
\tau = {{\sum_{i}{(\mathcal{R}_i-\mathcal{E}_i)}} \over {\sqrt{\sum_i{\mathcal{V}_i}}}}
\label{tauen}
\end{equation}
to determine the independence of two variables in a data set, say ($x_i,y_i$) for  $i=1, \dots, n$.  Here $R_i$ is the rank of variable $y$ of the data point $i$ in a set associated with it.  For a untruncated data (i.e. data truncated parallel to the axes) the {\it associated set} of point $i$ includes all of the data with  $x_j < x_i$.  If the data is truncated one must form the {\it associated set} consisting only of those points of lower $x$ value that would have been observed if they were at the $x$ value of point $i$ given the truncation, see definition below. 

If ($x_i,y_i$) were independent then the rank $\mathcal{R}_i$ should be distributed continuously between 0 and 1 with the expectation value $\mathcal{E}_i=(1/2)(i+1)$ and variance  $\mathcal{V}_i=(1/12)(i^{2}+1)$. Independence is rejected at the $n \, \sigma$ level if $\vert \, \tau \, \vert > n$.  
Here the mean and variance are calculated separately for each
associated set and summed accordingly to produce a single value for
$\tau$. This parameter represents the degree of correlation for the
entire sample with proper accounting for the data truncation.}

With this statistic, we find the parametrization that best describes the luminosity and time evolution.
This means that we have to determine the limiting flux, $F_{lim}$, which gives the minimum observed luminosity for a given redshift, $L_x=  4 \pi D_L^2(z) \, F_X K$ as shown in Fig. \ref{fig2}. The nominal limiting sensitivity of XRT, $F_{lim}=10^{-14}$ {\rm erg cm}$^{-2}$ ${\rm s}^{-1}$, is too low to describe the truncation of our sample, dashed line. This is because there is a limit in the plateau end times, $T^{*}_{a,{\rm lim}}= 242/(1+z)$ s, right panel of Fig. \ref{fig2}. Therefore, as pointed out by Cannizzo et al. 2011 this restriction increases the flux threshold to $10^{-12}$ erg cm$^{-2}$. Therefore, taking into account the above minimum plateau end time we have investigated several limiting fluxes to determine a good representative value while keeping an adequate size of the sample. We have chosen the limiting flux $F_{lim} = 1.5 \times $10$^{-12}$ erg cm$^{-2}$, shown by the red solid line, which allows 90 GRBs in the sample.

\subsection{Cosmological evolutions}

The first step required for this kind of investigation is the determination of whether
the variables $L_X$ and $T^*_a$, are correlated with redshift or are
statistically independent of it. For example, the correlation between $L_X$ and the redshift, $z$, is what we
call luminosity evolution, and independence of these variables
would imply absence of such evolution. 
The EP method prescribed how to remove the correlation by defining
new and independent variables.

Following the approach used for quasars and blazars \citep{Singal2011,Singal2012b,Singal2012a}, we determine the correlation functions, g(z) and f(z) when determining the evolution of $L_X$ and $T^{*}_a$  so that de-evolved variables $L'_{X} \equiv L_X/g(z)$ and $T'_a \equiv T^*_a/f(z)$ are not correlated with z. 
The evolutionary function are parametrized by simple correlation functions

\begin{equation}
g(z)=(1+z)^{k_{Lx}}, f(z)=(1+z)^{k_{T^{*}a}}
\label{lxev}
\end{equation}

so that $L'_{X}=L_X/g(z)$ refer to the local ($z=0$) luminosities. This is an arbitrary choice. One can chose any other fiducial redshift by defining $g(z)=[(1+z)/(1+z_{fid})]^{k_Lx}$. We have also tried this approach obtaining compatible results with the presented ones.
The associated set for the source $i$ to obtain the luminosity evolution is :
\begin{equation} \label{eq:Ji1}
    J_{i} \equiv \{j:z_{j} < z_{max}(L_i) \}  \vee  \{j:   L_j > L_i \},
   \end{equation}
 
where $z_{\rm max}(L_i)$ is the maximum redshift at which object $i$ with $L_j$ could be placed and still be included in the survey.
The objects of all the sample are indicated with $i$, while the objects in the associated sets are denoted with $j$.  With the the simbol $\vee$ we intend the union of the sets.

Analogously, to obtain the plateau end time evolution factor the associated set for a given object $i$ are :

\begin{equation} \label{eq:Ji2}
    J_{i} \equiv \{j:  z_j > z_{min,i} \}  \vee  \{j: T_j > T_i  \},
\end{equation}

\noindent where $z_{\rm min}(T_{a_i})$ is the minimum redshift at which object $i$ could be placed and still be included in the survey given its plateau duration and the limiting time of the observation.

\begin{figure}
\includegraphics[width=0.50\textwidth,height=0.48\textwidth]{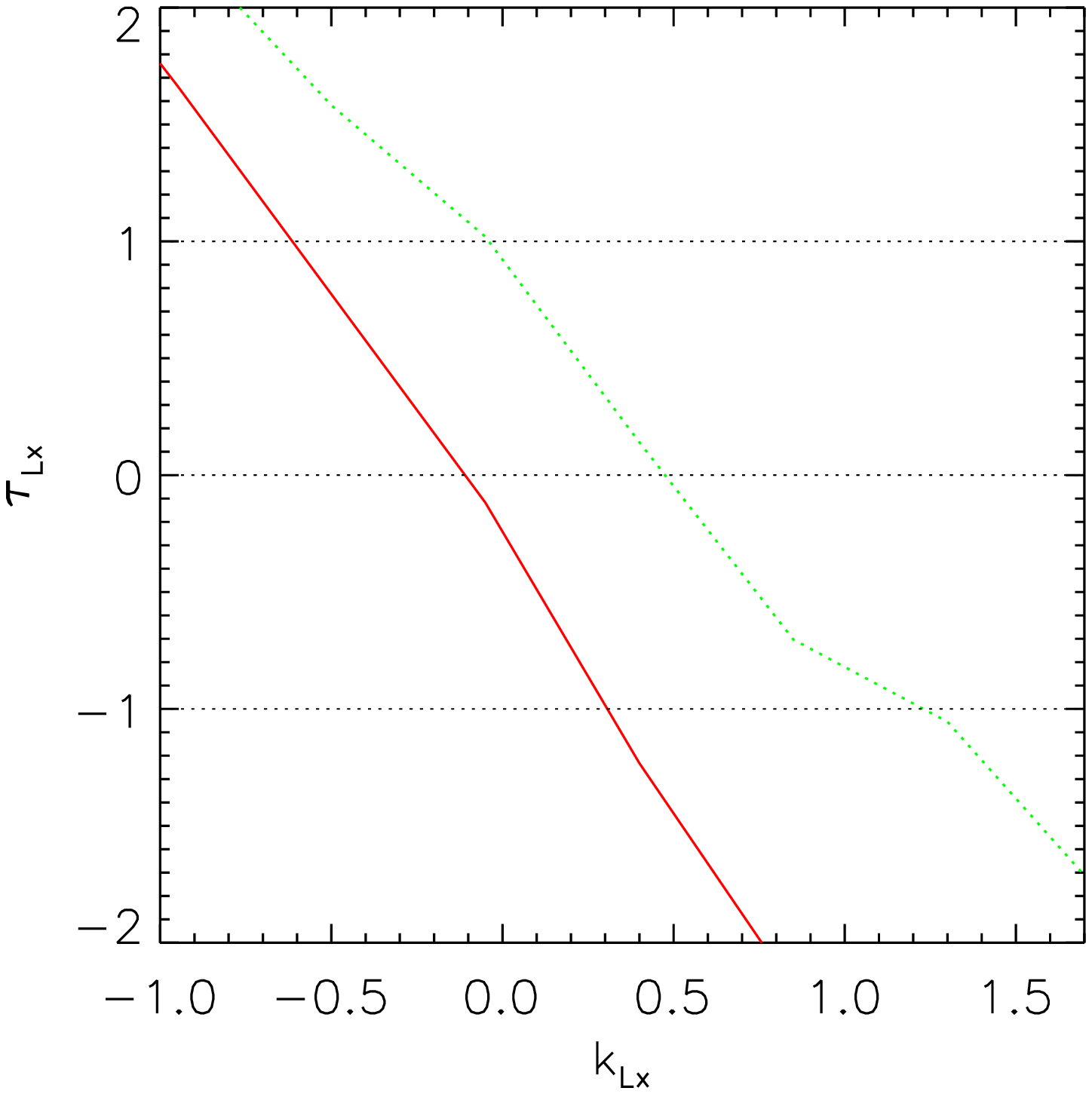}
\includegraphics[width=0.50\textwidth,height=0.50\textwidth]{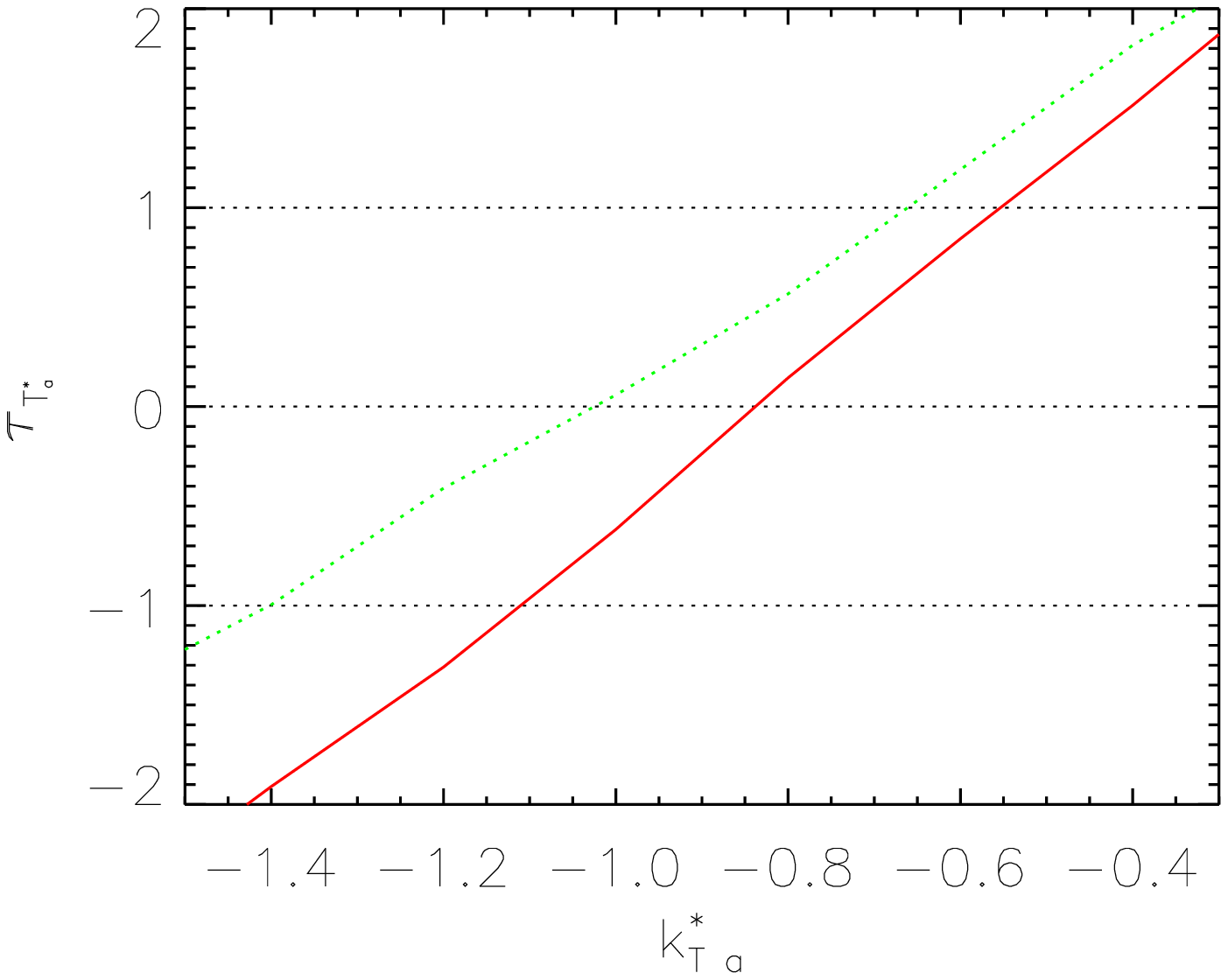}
\caption{ {\bf Left:} Test statistic $\tau$ vs. $k_{L_x}$, the luminosity evolution defined by Eq. \ref{eq:Ji1}. Right panel Test statistic $\tau$ vs. $k_{T^{*}_a}$, the time evolution defined by Eq. \ref{eq:Ji2}. The red line represents the full sample of 101 GRBs, while the green line represents the small sample of 47 GRBs in common with the previous sample of 77 GRBs.}
 \label{Fig4}
\end{figure}

With the specialized version of Kendell's $\tau$ statistic, the values of $k_{L_x}$ and $k_{T^{*}a}$ for which $\tau_{L_x} = 0$ and $\tau_{T^{*}a} = 0$ are the ones that best fit the luminosity and plateau end time evolution respectively, with the 1$\sigma$ range of uncertainty given by $| \tau_x | \leq 1$. Plots of $\tau_{L_x}$ and $\tau_{T^{*}a}$ versus $k_{L_x}$ and $\tau_{T^{*}a}$ are shown in Fig. \ref{Fig4}. With $k_{L_x}$ and $k_{T^{*}a}$ we are able to determine the de-evolved observables $T{'}_a$ and $L{'}_X$. 

We evident there is no discernable luminosity evolution, $k_{L_x}=-0.05_{-0.55}^{+0.35}$, but there is a significant evolution in $T^{*}_a$, $k_{T^{*}a}=-0.85_{-0.30}^{+0.30}$.
 
\subsection{Intrinsic LT correlation}

This is the first time the EP method has been applied in a parameter space for a bivariate correlation which involves a luminosity and a time, while previously the EP applications have been done in a luminosity-luminosity space \citep{Singal2012a}.
Thefore, we stress that this means different trend in the data truncation as we have shown in Fig \ref{fig2}. In the [$L_X-T^{*}_a$] variable space we apply the EP method to define the associated sets as:

\begin{equation} \label{eq:Ji}
    J_{i} \equiv \{j:  L^{'}_{\rm min}(z_j) < L^{'}(z_i)\} \vee \{ j: L^{'}_j > L^{'}_i \} \vee \{ j: T^{'}_{a_{min}}(z_j) < T^{'}{_a(z_i)} \},
\end{equation}
where $L^{'}_{\rm min}(z_j)$ and $T^{'}_{a_{min}}(z_j)$ are respectively the de-evolved minimum luminosity and de-evolved plateau time at redshift $z_j$ that object $j$ could have and still be included in the survey given the flux limits, its redshift, and the limiting time of the observation : 
$L_{\rm min}(z_j)=4 \pi D_L^2(z_j) \, F_{lim} \textit{K}$ and $T^{*}_{a_{min}}(z_j)=T_{a,{\rm lim}}/(1+z_j)$.
Using the Kendall $\tau$ rank test we determine whether $L_X$, $T^{*}_a$ are independent or not. The test shows some dependence, so we apply a coordinate transformation by defining a new luminosity $L^{'}_X$ as

\begin{equation} \label{eq:Ldeevolved}
\log_{10} L^{'}_X = \log Lx + \alpha \log T^{*}_a. 
\label{alphae}
\end{equation}
We then vary $\alpha$ and determine the value of $\tau$ (the correlation between $L^{'}_X$ and $T^{'}_a$) as a function of $\alpha$. The value of $\alpha$ which gives $\tau_{\alpha}=0$ determines the correlation between $L_X$ and $T^{*}_a$ with the 1$\sigma$ range of uncertainty given by $|\tau_{\alpha} | \leq 1$.

Fig. \ref{Fig3} shows variation of $\tau_{alpha}$ with $\alpha$. As can be seen independence is achieved ($\tau=0$) for $\alpha=1.07$ with a 1 $\sigma$ range of $\alpha=(-1.21,-0.98)$, therefore $\alpha=-1.07_{-0.14}^{+0.09}$. This means that $L_X$ and $T^{*}_a$ are correlated with the intrinsic slope of $1.07$ and that the significance of this correlation is at 12 $\sigma$ level. The $\alpha$ value is flatter than the one obtained from the raw data of the whole sample (parameter b) and it is compatible with the average value of the slopes of the subsamples shown in Table 1.
 
{\bf With the EP method we are able both to overcome the problem of selection effects and to determine the intrinsic value of the slope, because we removed the induced correlation by observables due to the time evolution and luminosity evolution dividing the respective time and luminosity for the respective evolution functions as it is explained above. Any differences between the correlation obtained from our methods and the present one in the raw data is assumed to arise from selection effects and partly to time evolution. The evolution seen in Figure 1 (central panel) is due to observational biases and partly to time evolution. In fact, we have determined that there is no cosmological evolution of $L_X$, and that the evolution of $T^{*}_a$ becomes significant at high redshift only.}

We also present results for a sample of 53 GRBs which are in common between a previous sample of 77 GRBs \citep{Dainotti2010} and the present one (see green line of Fig. \ref{Fig3} and Fig. \ref{Fig4}). For the sample of 53 GRBs we have adopted as a limiting flux $1.8 \times 10^{-12}$ erg cm$^{-2}$ s$^{-1}$ adopting the same criterion as for the larger sample. In this case we have 47 GRBs remaining above the adopted flux limit, again resulting in retain 90 $\%$ of the sample. We note that there is compatibility within 1$\sigma$ range among the power law slopes ($\alpha$) of the two samples. {\bf The two samples are fitted with a different fitting procedure, one procedure leaves $t_a$ free to vary (101 sample) and the other fixes $t_a=T_p$, where $T_p$ is the beginning of the decay phase of the prompt emission, therefore we here stress that we still find compatible results proving that the LT correlation intrinsic slope is independent from the particular adopted procedure.} 

In addition, for a consistency check we have used an inverted transformation:

\begin{equation}\label{eq:alphareciprocol}
\log_{10}(T^{'}_a) = \log T^{*}_a + \alpha^{*} \log L_X, 
\end{equation}
and followed the same procedure to determine $\tau$ as a function of $\alpha^{*}$. We expect $\alpha^{*}=1/\alpha$. The result is shown in the right panel of Fig. \ref{Fig.7}. Values of $\alpha^{*}=-0.71^{-0.10}_{+0.12}$ are compatible within 2 $\sigma$ with $1/\alpha=-0.93 \pm 0.09$ obtained from the previous transformation. This is a further demonstration that the method is well build and the results are robust. However, for an exact compatibility we would expect $\alpha^{*}=1/\alpha$ (See also the Appendix).

\section{The cumulative local luminosity and density functions}\label{density and luminosity}

Since we found no luminosity evolution the cumulative distribution of $\Phi(> L)=\int^{\infty}_{L} \Psi (L^{'}) dL^{'}$ according to our method \citep{Petrosian1992} is given as:

\begin{equation}
\Phi_i\!(L_{i}) = \prod_{k}{(1 + {1 \over n\!(k)})}
\label{phieq}
\end{equation}

Here $n(k)$ is the number of objects in the associated sets of object k, namely these with $L>L_k$ and $z<z_{max}$.

The density rate evolution $\rho\!(z)^{'}$ and the LF, (with ${'}$ we indicate the differential simbol), which gives the number of objects per unit comoving volume $V$ per unit source luminosity can be computed with the EP method.
The method gives the cumulative functions $\sigma(<z)^{'}=\int_0^z{\rho}(z')\,[dV(z')/dz']\,dz'$ and $\phi(>L')=\int_{L'}^\infty \psi(L'')\,dL''$. The differential functions $\rho^{'}$ and $\psi$ are obtained by differentiation.

\begin{figure}
\includegraphics[width=0.50\hsize,angle=0,clip]{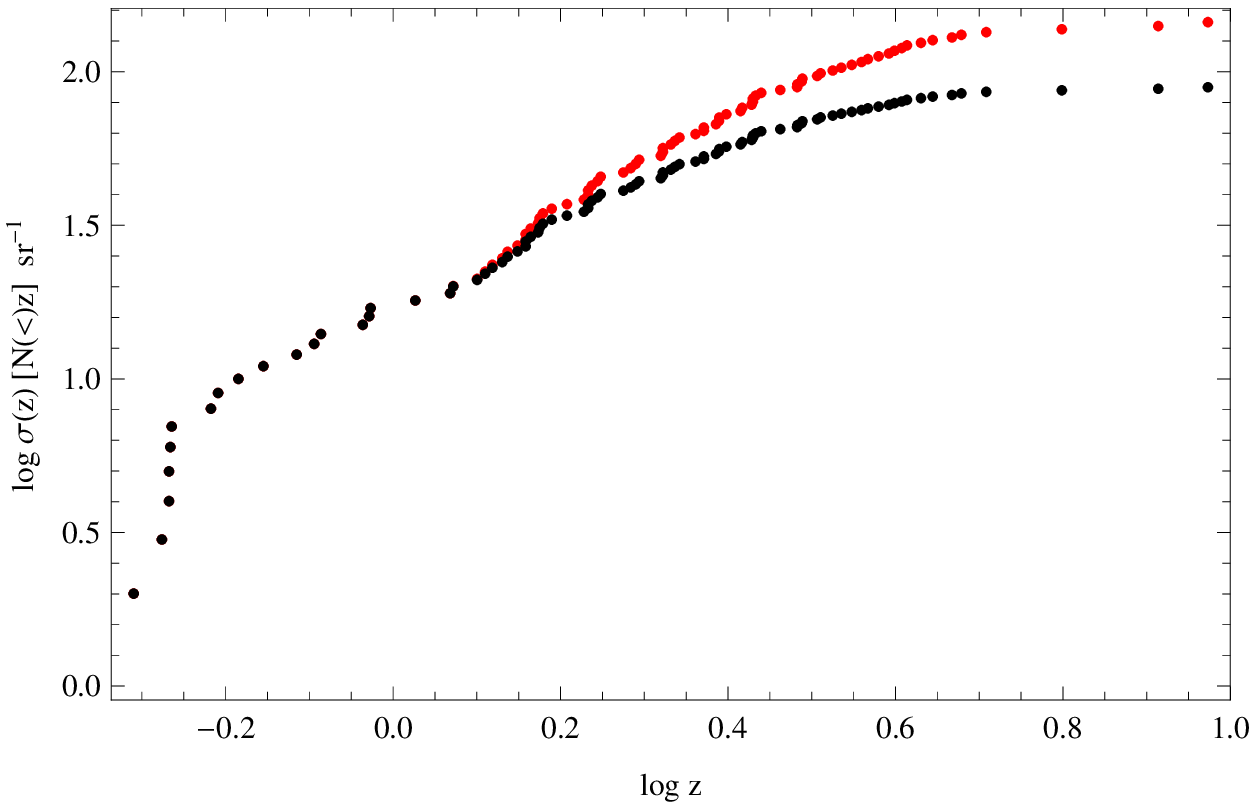}
\includegraphics[width=0.50\hsize,angle=0,clip]{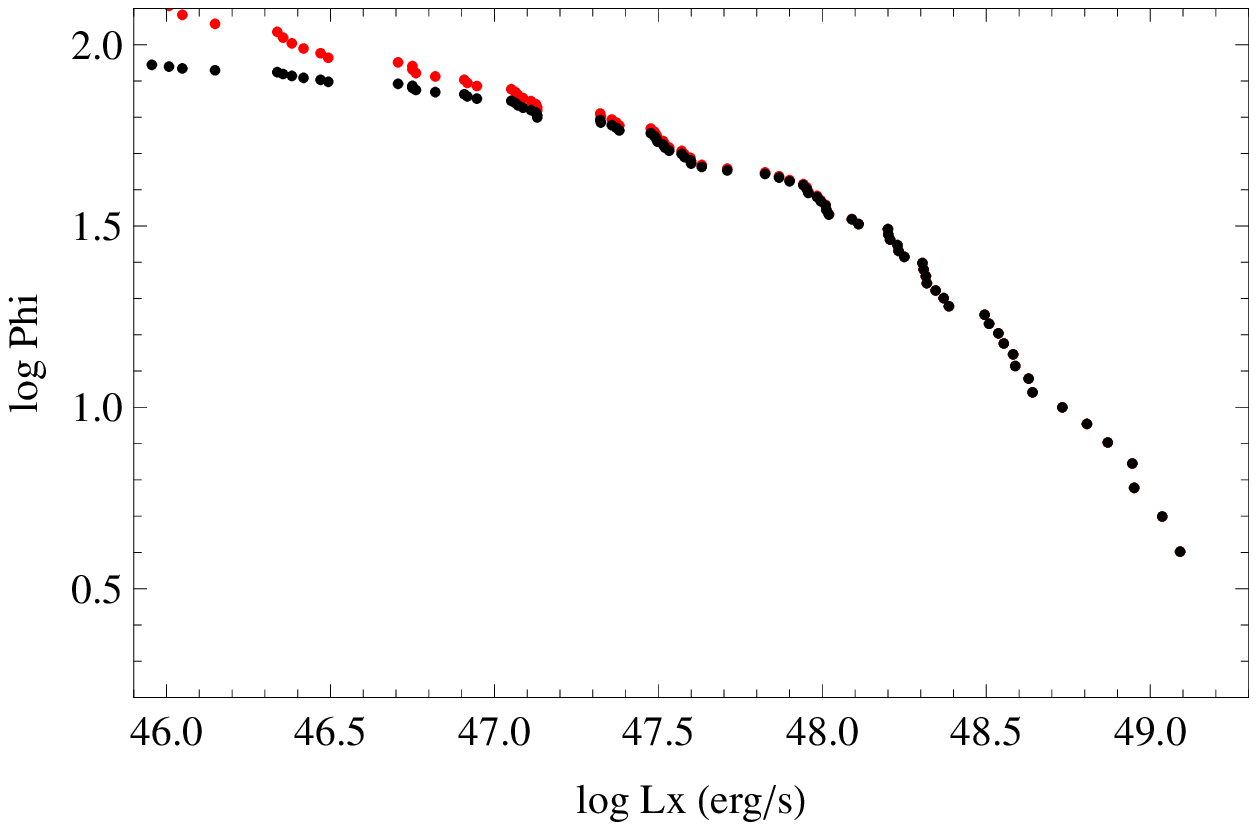}
\caption{ {\bf Left panel} : cumulative redshift distribution, N(>z) of the raw data (black points) and the cumulative distribution, $\sigma(z)$ corrected by the EP method (red points) and described in Section \ref{density and luminosity}.{\bf Right panel}: the cumulative local luminosity distribution of raw data (black point) and corrected by the EP method (red points).}
 \label{Fig.5}
\end{figure}

One can define the cumulative density function as follow:

\begin{equation}
\sigma\!(z_j) = \prod_{i}{(1 + {1 \over m\!(i)})}
\end{equation}

where $i$ runs over all objects with a redshift lower than or equal to $z$, and $m(i)$ is the number of objects with a redshift lower than the redshift of object $i$ {\it which are in object j's associated set}.  In this case, the associated set is again those objects with X-ray luminosity that would be seen if they were at object $i$'s redshift. The use of only the associated set for each object removes the biases introduced by the data truncation.  
We show the distribution in the right panel of Fig. \ref{Fig.5} of the cumulative density distribution corrected (red points), which is contrasted with the raw distribution (black points).

As evident the correction for the cumulative density starts to apply for $z=1$, namely we have a higher density of GRBs than the one we observe for $z>1$.
In Fig. \ref{Fig.5} (left panel) the corrected cumulative luminosity function agrees with the raw observed luminosity distribution until $L_{X}=10^{48}$ erg/s while for higher values of the luminosity the two distributions separate.

\begin{figure}
\includegraphics[width=0.50\hsize,angle=0,clip]{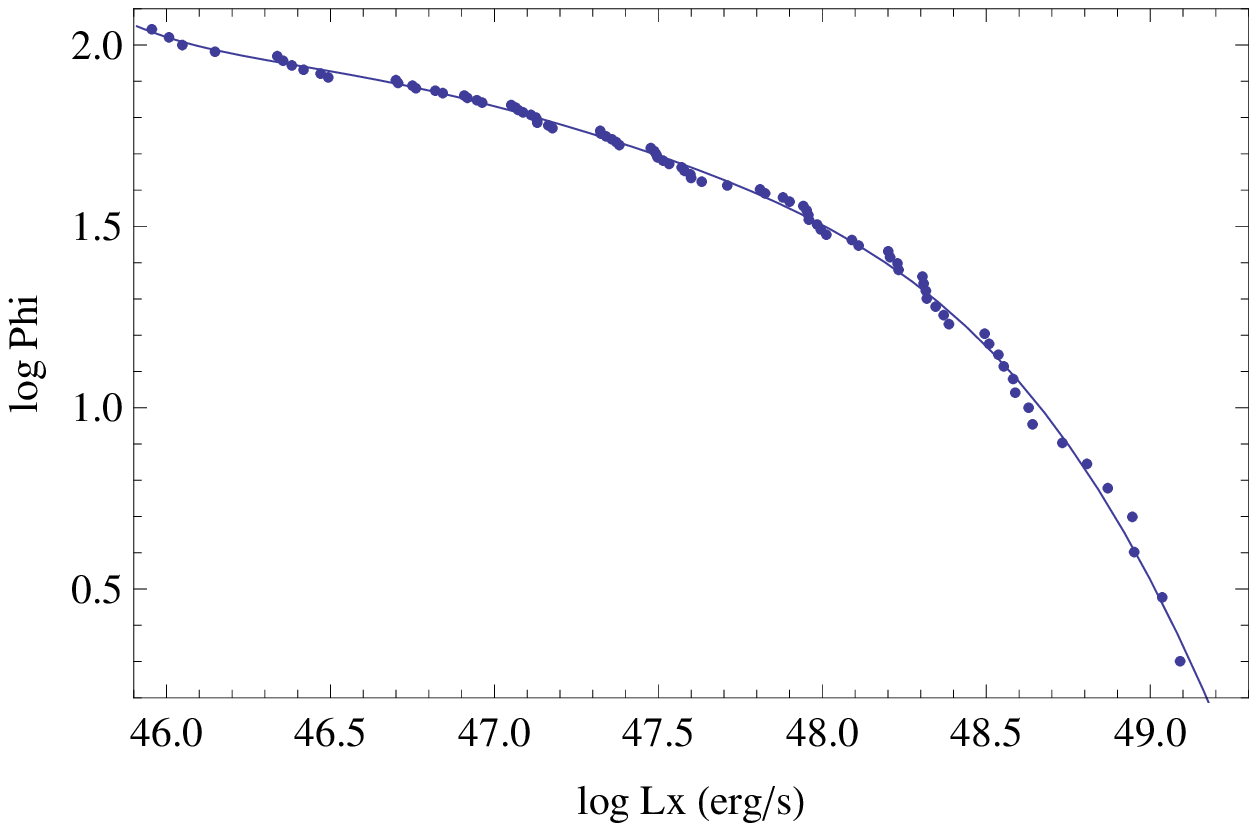}
\includegraphics[width=0.50\hsize,angle=0,clip]{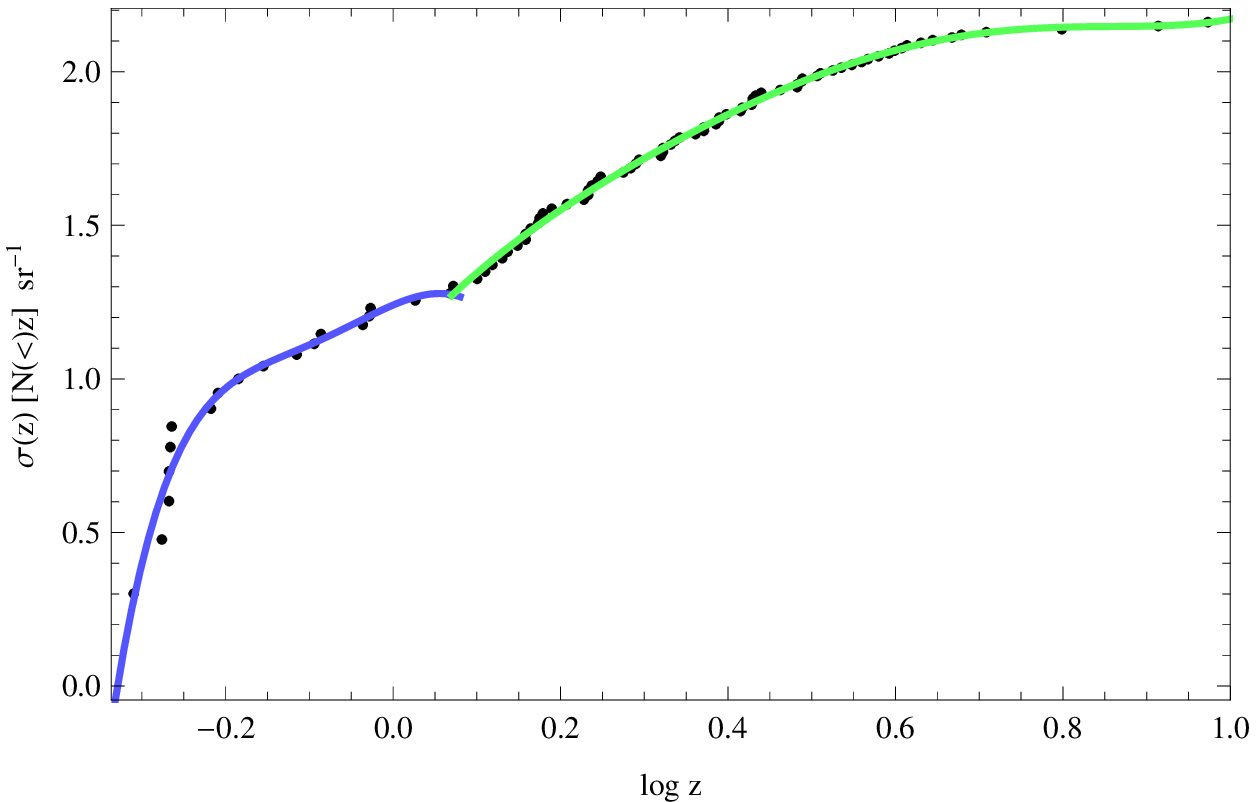}
\caption{{\bf Left Panel }: cumulative intrinsic luminosity function determined by the EP method along with a fitted function as discussed in Section 
\ref{density and luminosity}. {\bf Right panel }: the cumulative intrinsic density distribution with two fitted functions lines, as discussed in Section 
\ref{density and luminosity}. The blue line till $z=1$, while the green line for $z>1$.}
 \label{Fig.6}
\end{figure}

To obtain the differential distribution $\psi(L)$ and $\rho^{'}(z)$ we fitted the cumulative luminosity function with a polinomial of order $7$, while for the cumulative density we divide the distribution in two parts, one with $z \leq 1$ as we can see from Fig. \ref{Fig.6}, blue line and the other part for $z \geq 1$, green line. The two fitting lines are both a polinomial of order $5$.

\section{Discussion}

The obtained $\tau_{\alpha}$ vs $\alpha$ plot (Right panel of Figure \ref{Fig3}), clearly demonstrates the existence at the 12 $\sigma$ level of a significant LT correlation, characterized by the value found for the power law slope relating the luminosity and the plateau end times.

Therefore, the presented analysis, with the intrinsic value of the power law slope of the LT correlation, provides new constraints for physical models of GRB explosion mechanisms. 
With this new determination of the correlation power law slope we discuss the consequences of these findings for GRB physical models. The LT relation is predicted by several theoretical models \citep{Cannizzo09,Cannizzo11,Dall'Osso,Yamazaki09,Obrien2012,Bernardini2011} and in other observational ones \citep{Ghisellini2009,Qi2012}, proposed for the physical GRB evolution in the time $T_a$. Recently, Oates et al. 2012 pointed out the existence of an anticorrelation between the luminosity at $200$ s and the decay slope of the optical lightcurve. This correlation is related to the LT one considered here. Racusin et al. (in prep.) recover the Oates et al. correlation in the X-ray band. Therefore, it is now even more challenging to understand the meaning of the $L_X$ and $T^{*}_a$ correlation, which becomes the principal X-ray afterglow correlation from which further correlations in other wavelengths can be derived. From a theoretical point of view the Cannizzo model predicts a correlation slope ($3/2$) which is in agreement with our intrinsic correlation power law only within $3\sigma$, while the model of Yamazaki predicts a less steep decay which is in agreement in 1$\sigma$ with the presented results. The LT correlation is also recovered for short GRBs \citep{Obrien2012} within the magnetar scenario. Any physical interpretation of the LT correlation should be based on the intrinsic power slope and not that obtained from the raw observed quantities. In fact, assuming the observed power law as a key feature to discriminate among physical models could lead to misleading results based either on data truncation or on redshift evolution. 
We conclude that determining the intrinsic correlations among, and distributions of, the observables is a necessary step before any possible and plausible usage of the LT correlation as a theoretical model discriminator, distance estimator, and as useful cosmological tool. Therefore, this paper opens a new perspective not only on the interpretation of the LT correlation but also of the other existing GRB correlations and prepares for a new possible future approach for the usage of GRBs in cosmology.

\section{Appendix}

\begin{figure}
\includegraphics[width=0.33\hsize,angle=0,clip]{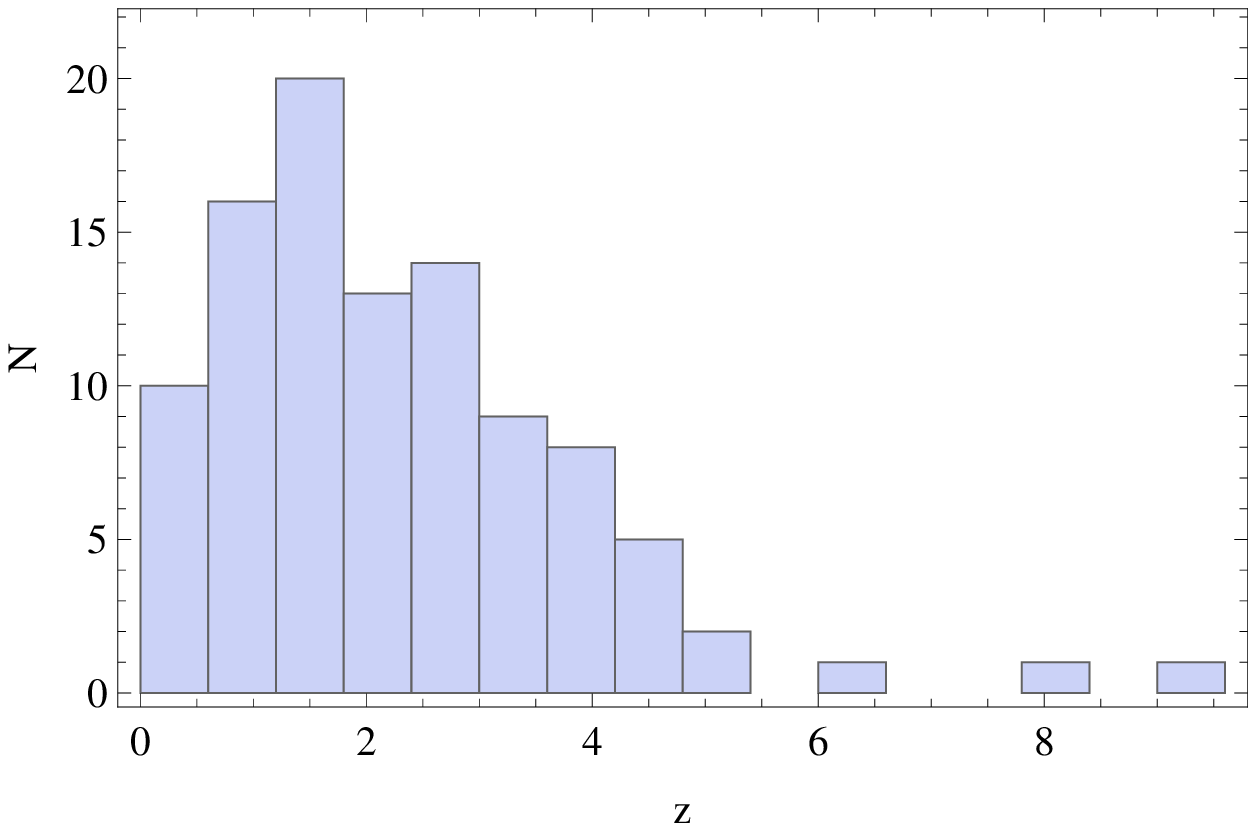}
\includegraphics[width=0.32\hsize,angle=0,clip]{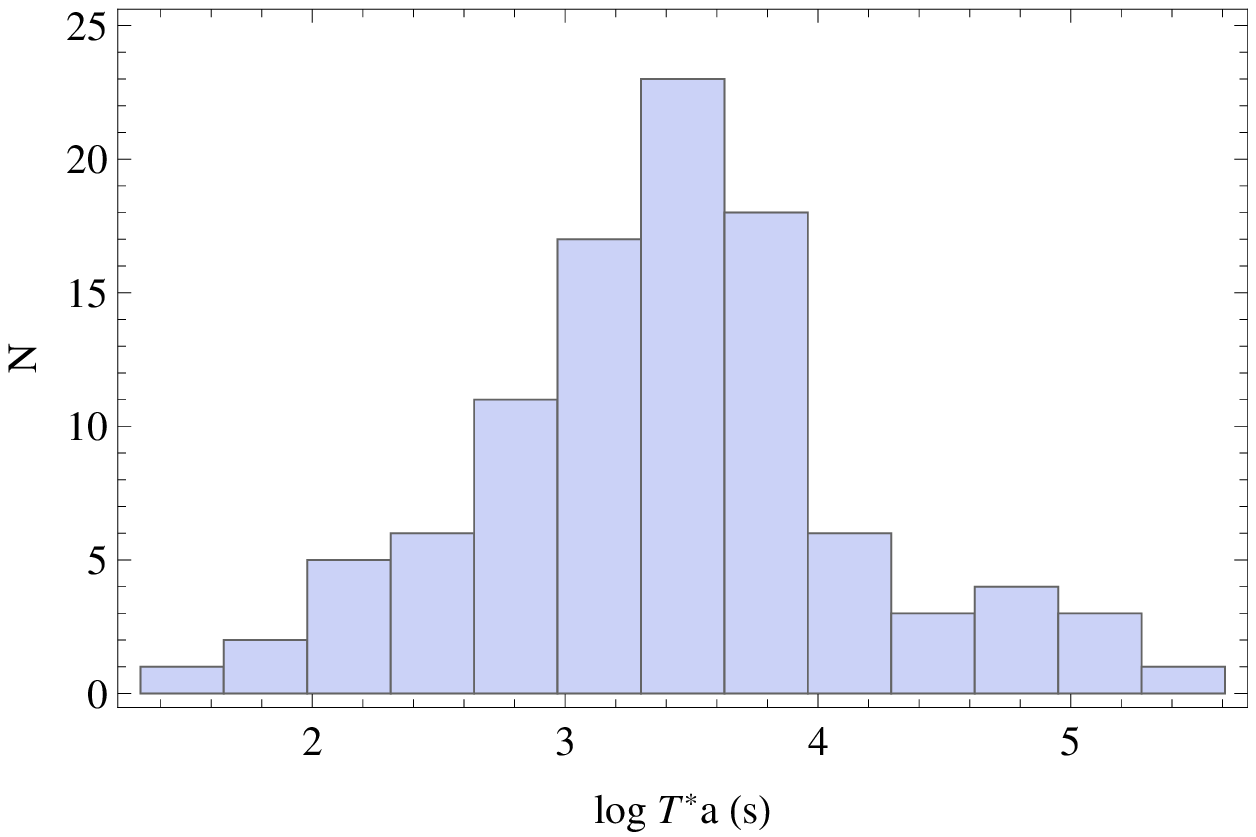}
\includegraphics[width=0.33\hsize,angle=0,clip]{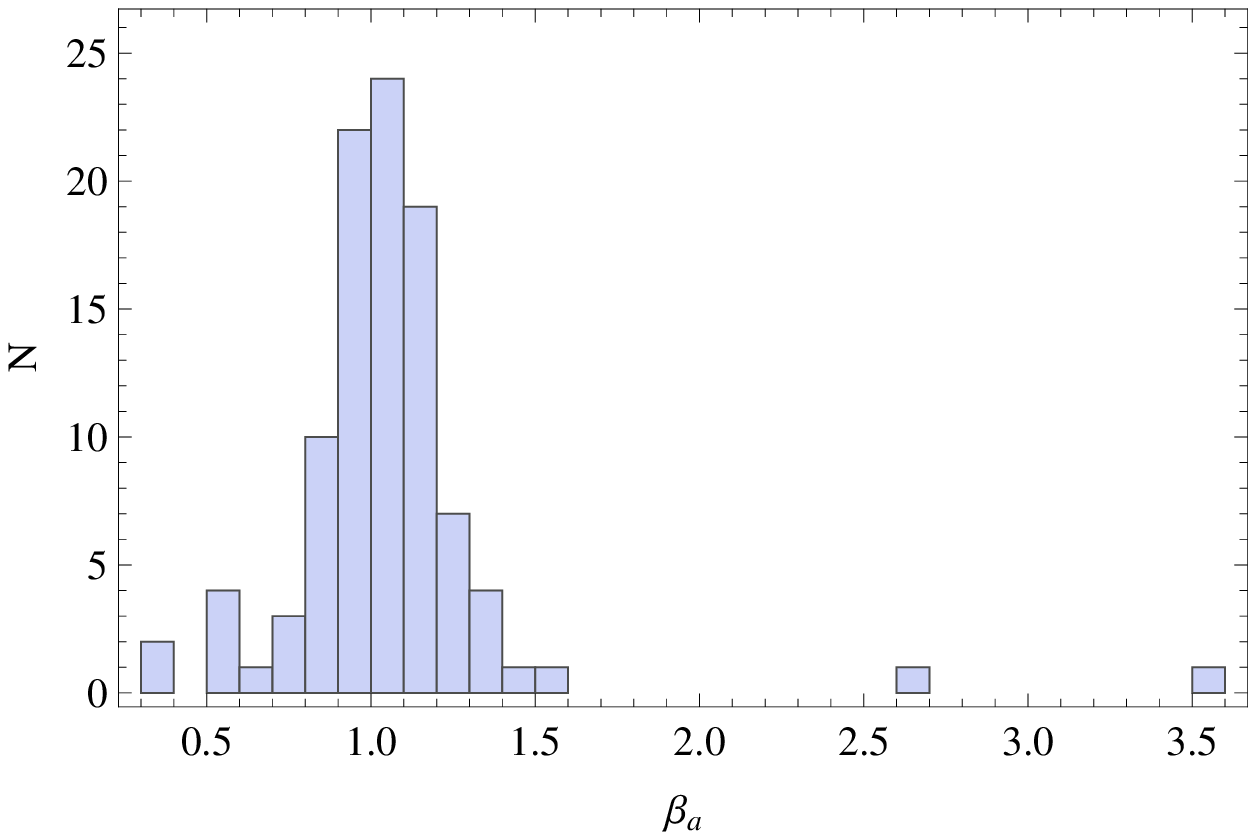}
\includegraphics[width=0.33\hsize,angle=0,clip]{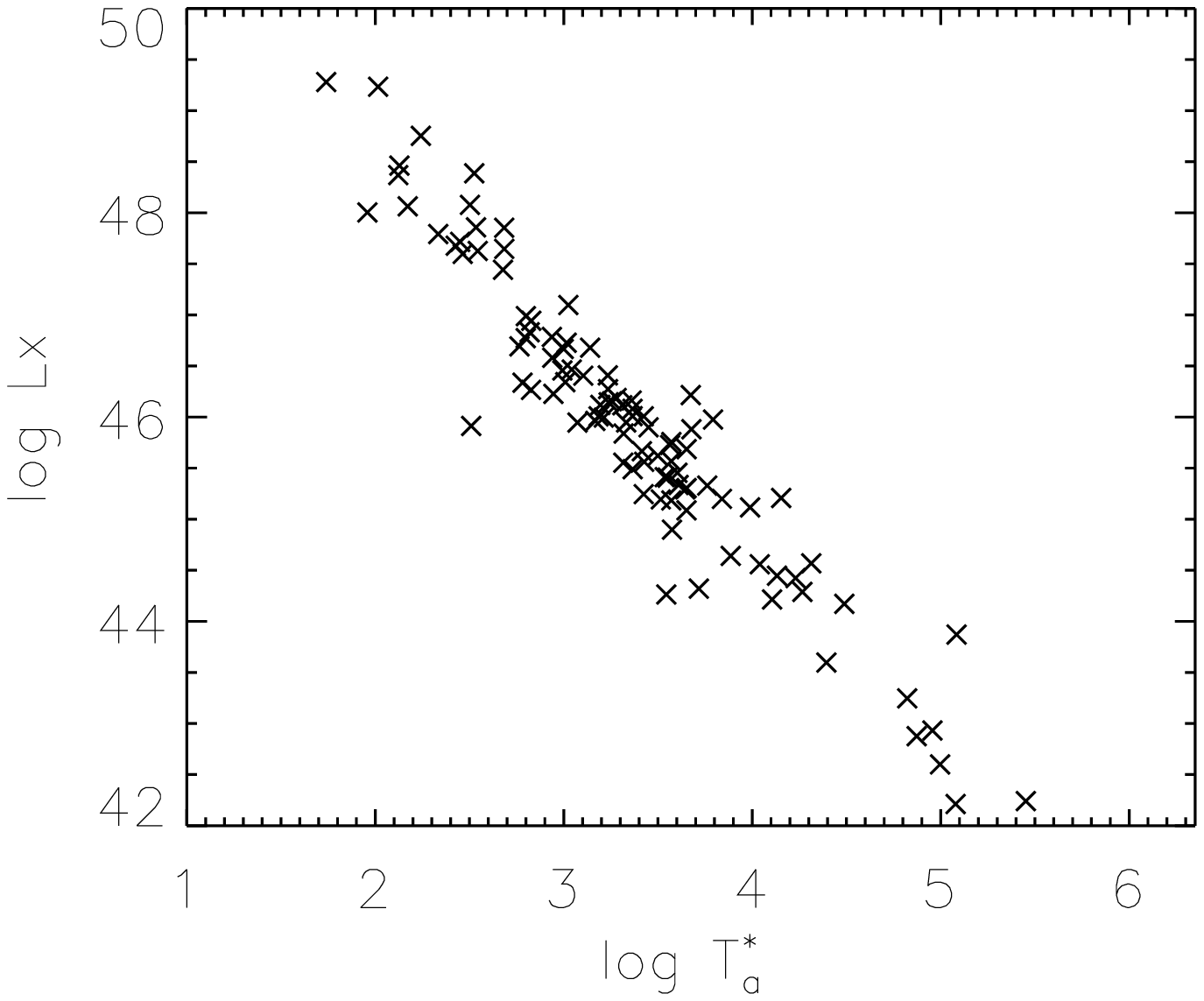}
\includegraphics[width=0.32\hsize,angle=0,clip]{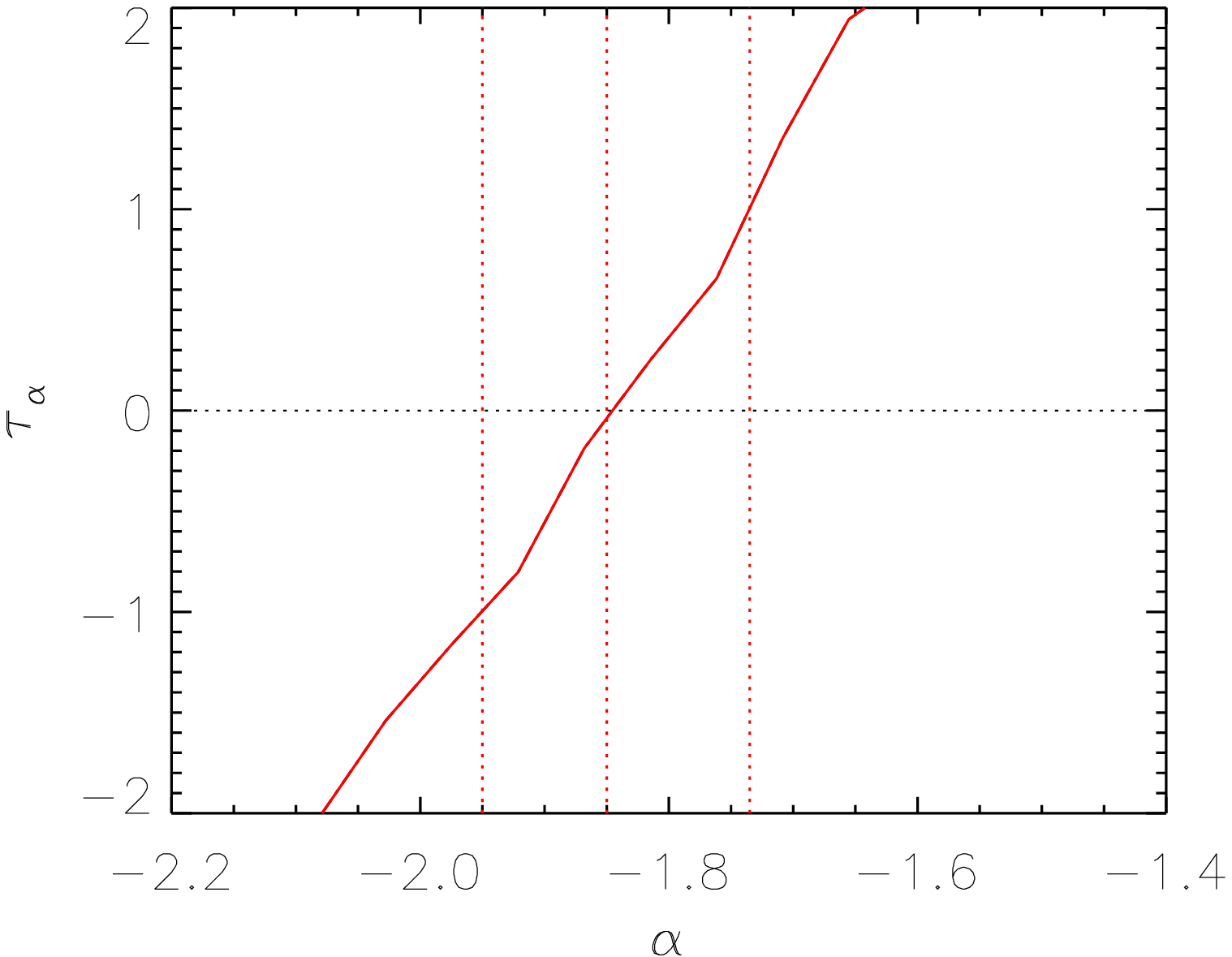}
\includegraphics[width=0.33\hsize,angle=0,clip]{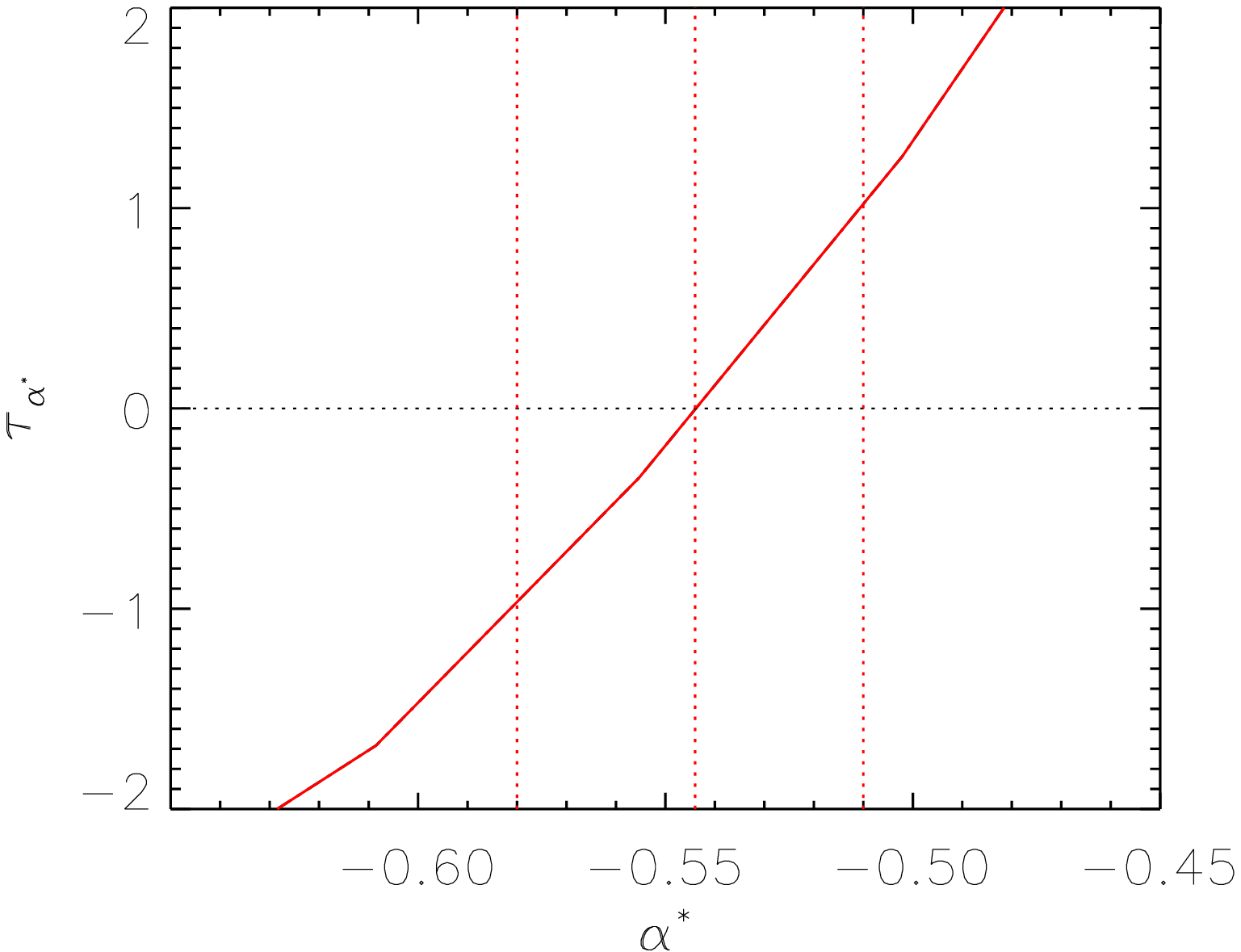}
\caption{ {\bf Upper Panel}: The distribution of the redshift (left), of the time $T^{*}_a$ (middle) and of the spectral index (right) for the sample of 101 GRBs used in the analysis.
 {\bf Lower panel}: $L_X$-$T^*_a$ distribution for a set of 101 simulated GRBs discussed in Appendix (left); Test statistic $\tau$ vs. the LT correlation power slope parameter $\alpha$ (center), and the reciprocol of it, $\alpha^{*}$, (right) for the simulated set. The 1$\sigma$ range of best fit values is where $| \tau | \leq 1$ shown by the vertical dotted lines.}
\label{Fig.7}
\end{figure}

For a further test of the robustness of the main conclusion of this work we have applied the analysis methods discussed here to a simulated observational data set with a known intrinsic LT correlation. As is clear from the top panel of \ref{Fig.7} the distributions of three observables in the real data,  the time, $T^{*}_a$, the spectral index, $\beta_a$,and the redshift, $z$, can be approximated with normal distributions with mean values which are $<\beta_a>=1.05$, $<T^{*}_a>=3.5$ and $<z>=2.09$. Therefore, we have created a Monte Carlo population with these distributions. The luminosities are determined by applying an LT correlation with $\log L_X \approx -1.9 \log T^{*}_a$, $-1.9$ in this case being  the imposed $\alpha$ slope of LT correlation (see left lower panel Fig. \ref{Fig.7}). We then compute the simulated fluxes from the simulated $\beta$, $z$ and $L_X$. We have imposed the same limiting flux used for the real observational data to form an `observed' simulated data set on which we then apply the analysis method discussed in this work. Application of the method successfully recover the known intrinsic power law slope of the LT correlation and its inverse as is shown in lower central and right panels of Fig. \ref{Fig.7}.

\section{Acknowledgments}
This work made use of data supplied by the UK Swift Science Data Centre at the University of Leicester. M.D. is grateful to Richard Willingale and Paul Obrien for comments on the paper and Qin Rong Chen for fruitful discussions. M.D is also grateful to the Polish MNiSW through the grant N N203 579840, the Fulbright Scholarship and the Ludovisi- Blanceflor Foundation.  M. O. is grateful to the Polish National Science Centre through the grant DEC-2012/04/A/ST9/00083.

\end{document}